\documentclass[prd,nofootinbib,superscriptaddress,eqsecnum,showpacs]{revtex4}
\def\G{\Gamma}
\def\H{{\cal H}}
\def\w{\Omega}
\def\C{{\bf C}}
\def\c{{\cal C}}
\def\S{{\cal S}}
\def\F{{\cal F}}
\def\la{\langle}
\def\ra{\rangle}
\def\be{\nopagebreak[3]\begin{equation}}
        \def\ee{\end{equation}}
        \def\ba{\nopagebreak[3]\begin{eqnarray}}
        \def\ea{\end{eqnarray}}
        \def\nl{\nonumber \\}
        
\def\d{{\rm d}}

\newcommand{\teta}{\rlap{\lower2ex\hbox{$\,\tilde{}$}}\eta{}}

\newcommand{\ld}{N_{_{\!\!\!\!\!\sim}}}
\begin{document}

\preprint{\vbox{\baselineskip=12pt \rightline{ICN-UNAM-02/01}
\rightline{hep-th/0202070} }}

\title{On the Relation Between Fock and\\
Schr\"odinger Representations for a Scalar Field}
\author{Alejandro Corichi}\email{corichi@nuclecu.unam.mx}
\affiliation{Instituto de Ciencias Nucleares\\
Universidad Nacional Aut\'onoma de M\'exico\\
A. Postal 70-543, M\'exico D.F. 04510, MEXICO \\
}
\affiliation{Department of Physics and Astronomy\\
University of Mississippi, University, MS 38677, USA\\
} \author{Jer\'onimo Cortez}\email{cortez@nuclecu.unam.mx}
\affiliation{Instituto de Ciencias Nucleares\\
Universidad Nacional Aut\'onoma de M\'exico\\
A. Postal 70-543, M\'exico D.F. 04510, MEXICO \\
}

\author{Hernando Quevedo}\email{quevedo@physics.ucdavis.edu}
\affiliation{Instituto de Ciencias Nucleares\\
Universidad Nacional Aut\'onoma de M\'exico\\
A. Postal 70-543, M\'exico D.F. 04510, MEXICO \\
}
\affiliation{Department of Physics\\
University of California\\
Davis, CA 95616, USA\\
}

\begin{abstract}
Linear  free field theories are one of the few Quantum Field
Theories that are exactly soluble. There are, however, (at least)
two very different languages to describe them, Fock space methods
and the Schr\"odinger functional description.  In this paper, the
precise sense in which the two representations are related is
reviewed. Several properties of these representations are studied,
among them the well known fact that the Schr\"odinger counterpart
of the usual Fock representation is described by a Gaussian
measure. A real scalar field theory is considered, both on
Minkowski spacetime for arbitrary, non-inertial embeddings of the
Cauchy surface, and for arbitrary (globally hyperbolic) curved
spacetimes. As a concrete example, the Schr\"odinger
representation on stationary and homogeneous cosmological
spacetimes is constructed.

\end{abstract}
 \maketitle

\section{Introduction}
\label{sec:1}

In ordinary quantum theory, a state is represented by a vector
$|\Psi\rangle$ in an abstract Hilbert space $\H$, and observables
are represented by Hermitian operators on $\H$. For a finite
dimensional system --such as a harmonic oscillator-- with a
configuration space with coordinates $q^i$, there are at least two
equivalent representations of the Hilbert space. One can represent
the quantum states as wave functions $\psi(q^i)=\langle
q^i|\Psi\rangle$ of the configuration space, in what is known as
the Schr\"odinger picture; alternatively, one can consider the
basis $\langle \vec{n}|$  such that a state is given by $\langle
\vec{n}|\Psi\rangle$, where the kets $|\vec{n}\rangle$ are
eigen-kets of the Hamiltonian:
$\hat{H}|\vec{n}\rangle=E(\vec{n})|\vec{n}\rangle$. In this case,
the state is characterized by the countable array of numbers
$\langle \vec{n}|\Psi\rangle$. Given the basis $|\vec{n}\rangle$,
and a general state
$|\Psi\rangle=\sum_{\vec{n}}A_{\vec{n}}|\vec{n}\rangle$ one can
then write  $\langle
q^i|\Psi\rangle\sum_{\vec{n}}A_{\vec{n}}\langle
q^i|\vec{n}\rangle$. Using the relation $\langle
q^i|\Psi\rangle=\sum_{\vec{n}}\langle
q^i|\vec{n}\rangle\langle\vec{n}|\Psi\rangle$, we conclude that
$A_{\vec{n}}=\langle\vec{n}|\Psi\rangle$. When we have knowledge
of the eigenfunctions as in the case of a harmonic oscillator
where $\langle q^i|\vec{n}\rangle=H_{\vec{n}}(q^i)$, we have a
natural passage from the `Fock' representation
$\langle\vec{n}|\Psi\rangle$ to the Schr\"odinger wavefunction
$\langle q^i|\Psi\rangle$. In this case both representations are
well defined and understood. A natural generalization of these two
descriptions for mechanical systems is to go to field theory. In
this case, the situation becomes more involved, due to the
existence of an infinite number of degrees of freedom.  In the
ordinary presentation of field theory, it is the Fock
representation the one that is more intuitive and widely known,
whereas the Schr\"odinger is not equally treated. This paper has
as its main motivation the desire of presenting, in a clear
fashion, both representations for a scalar field theory and their
relation.

The reason to study the quantum theory of a free real scalar field
is that it is one of the simplest field theory systems. Indeed, it
is studied in the first chapters on most field theory textbooks
\cite{textbook}. The language used for these treatments normally
involves Fourier decomposition of the field and creation and
annihilation operators associated with an infinite chain of
harmonic oscillators. Canonical quantization is normally performed
by representing these operators on Fock space  and implementing
the Hamiltonian operator. On the other hand, books that introduce
QFT from an axiomatic viewpoint \cite{cpt}, normally deal either
with functional {\it Euclidean} methods \cite{glimm} or with
abstract algebras and states in the algebraic approach
\cite{haag,baez}. All these approaches deal with QFT on a flat
Minkowski spacetime.
 An intermediate
treatment, motivated by the  the process of quantization, starting
from a classical algebra of observables and constructing
representations of them on Hilbert spaces \cite{wald2}, studies
quantum fields on curved spacetimes. This approach, closely
related to the classical ``canonical quantization" methods of
Dirac, is somewhat complementary to the standard viewpoint
\cite{birrell}. In the mentioned book, Wald develops the quantum
theory of a scalar field, and extends the formalism to an
arbitrary curved manifold. His construction is, however,
restricted to finding a representation on Fock space, or as is
normally known, the {\it Fock representation} \cite{wald2}.

On the other side of the program of canonical quantization for
fields, is the Schr\"odinger representation, where a slicing of
the spacetime is normally introduced (for reviews see
\cite{reviews}). This functional viewpoint, even when popular in
the past, is not widely used, in particular since it is not the
most convenient one for performing calculations of physical
scattering processes in ordinary QFT.\footnote{However, it has
been successfully used for proving a variety of results that do
not need dynamical information \cite{jackiw}.}
 However, from the conceptual
viewpoint, the study of the Schr\"odinger representation in field
theory is extremely important and has not been, from our viewpoint
widely acknowledged (however, see \cite{jackiw}).
 This is specially true since some symmetry reduced gravitational systems
can be rewritten as the theory of a scalar field on a {\it
fiducial, flat}, background manifold. In particular, of recent
interest are the polarized Einstein-Rosen waves \cite{einstein}
and Gowdy cosmologies \cite{gowdy}. In this regard, the
Schr\"odinger picture is, in a sense, the most natural
representation from the viewpoint of canonical quantum gravity,
where one starts from the outset with a decomposition of spacetime
into a spatial manifold $\Sigma$ ``evolving in time". Therefore,
it is extremely important to have a good understanding of the
mathematical constructs behind this representation and its
relation to the Fock representation.

The purpose of this paper is twofold. The first one is to review
and present the relation between the Fock and Schr\"odinger
representations of a scalar field on Minkowski spacetime, in a
coherent and unified manner.
We recall basic constructions at both  the classical and quantum
levels, and develop from a logical viewpoint the precise sense in
which the representations are related. In particular, we show in
detail the way in which the quantum measure in the functional
picture arises and its Gaussian character. In the process, we
emphasize the relevant geometrical objects that need to be
specified in order to define both representations and its
relation. In this regard, the present work can be considered as a
review. The emphasis in our presentation regarding this extra
structure allows us to achieve the second purpose of the paper,
namely, to extend the formalism to arbitrary embeddings of the
Cauchy surface and to arbitrary curved spacetimes, in the spirit
of \cite{wald2}. The generalization is at two levels. To be
specific, both the existing ambiguity in the quantization of a
scalar field, and the infinite freedom  in the choice of embedding
of the Cauchy surface are considered. We construct the relation
between these two representations and give concrete examples for
background spacetimes that are of physical interest. These later
results are, to the best of our knowledge, rather new and not
widely available.

 The structure of the paper is as follows.
In Sec.~\ref{sec:2} we recall basic notions from canonical
quantization and the classical formulation of a scalar field. In
Sec.~\ref{sec:3} we recall the ordinary Fock representation in the
spirit of \cite{wald2}. A discussion of the Schr\"odinger
representation and construction of the theory unitary equivalent
to a given Fock representation is the subject of
Section~\ref{sec:4}.  In Sec.~\ref{sec:5} we show the  relation
between the two equivalent representations in an explicit fashion.
In Sec.~\ref{sec:6} we present several examples of spacetimes for
which the Schr\"odinger representation is explicitly constructed.
We end with a discussion in Sec.~\ref{sec:7}. In a series of
appendixes, we prove some of the results presented in the main
text.

 In order to make this work accessible not
only to specialized researchers in theoretical physics, we have
intentionally avoided going into details regarding functional
analytic issues and other mathematically sophisticated
constructions. Instead, we refer to the specialized literature and
use those results in a less sophisticated way, emphasizing at each
step their physical significance. This allows us to present our
results in a self-contained fashion.

Throughout the paper we shall use units such that $G=c=1$ and the
abstract index notation for tensorial objects \cite{wald1}.

\section{Preliminaries}
\label{sec:2}

In this section we shall present some background material, both in
classical and quantum mechanics. This section has two parts. In
the first one we recall some basic notions of symplectic geometry
that play a fundamental role in the Hamiltonian description of
classical systems, and outline the canonical quantization starting
from a classical system. In the second part, we recall the phase
space description for a scalar field.

\subsection{Canonical Quantization}
\label{sec:2.a}

A physical system is normally represented, at the classical level,
by a {\it phase space}, consisting of a manifold  $\Gamma$ of even
dimension. The symplectic two-form $\Omega$ endows it with the
structure of a symplectic space $(\Gamma,\Omega)$. The symplectic
structure $\Omega$ defines the Poisson bracket $\{\cdot,\cdot\}$
on the observables, that is, on functions $f,g:\Gamma\rightarrow
R$, in the usual way: $\{f,g\}=\Omega^{ab}\nabla_a f\nabla_b g$.

In very broad terms, by quantization one means the passage from a
classical system, to a quantum system. Observables on $\Gamma$ are
to be promoted to self-adjoint operators on a Hilbert space.
However, we know that not all observables can be promoted
unambiguously to quantum operators satisfying the Canonical
Commutation Relations (CCR). A well known example of such problem
is factor ordering. What we {\it can} do is to construct a subset
${\cal S}$ of {\it elementary classical variables} for which the
quantization process has no ambiguity. This set ${\cal S}$ should
satisfy two properties:
\begin{itemize}
\item ${\cal S}$ should be a vector space large enough so that every
(regular) function on $\Gamma$ can be obtained by (possibly a
limit of) sums of products of elements in ${\cal S}$. The purpose
of this condition is that we want that enough observables are to
be unambiguously quantized.

\item The set ${\cal S}$ should be small enough such that it is closed
under Poisson brackets.

\end{itemize}

The next step is to construct an (abstract) quantum algebra ${\cal
A}$ of observables from the vector space ${\cal S}$ as the free
associative algebra generated by ${\cal S}$ (for a definition and
discussion of free associative algebras see \cite{geroch2}). It is
in this quantum algebra ${\cal A}$ that we impose the Dirac
quantization condition: Given $A,B$ and $\{A,B\}$ in ${\cal S}$ we
impose, \be
[\hat{A},\hat{B}]=i\hbar\widehat{\{A,B\}}\label{diracc} \, .\ee It
is important to note that there is no factor order ambiguity in
the Dirac condition since $A,B$ and $\{A,B\}$ are contained in
${\cal S}$ and they have associated a unique element of ${\cal
A}$.

The last step is to find a Hilbert space $\H$ and a representation
of the elements of ${\cal A}$ as operators on $\H$. For details of
this approach to quantization see \cite{tate}.

In the case that the phase space $\Gamma$ is a linear space, there
is a particular simple choice for the set $\S$. We can take a
global chart on $\Gamma$ and we can choose $\S$ to be the vector
space generated
 by {\it linear} functions on $\Gamma$. In some sense this is the smallest
choice of $\S$ one can take. As a concrete case, let us look at
the example
 of
$\c= R^3$. We can take a global chart on $\Gamma$ given by
$(q^i,p_i)$ and consider
$\S=\mbox{Span}\{1,q^1,q^2,q^3,p_1,p_2,p_3\}$. It is a seven
dimensional vector space. Notice that we have included the
constant functions on $\Gamma$, generated by the unit function
since we know that $\{q^i,p_i\}=1$, and we want $\S$ to be closed
under Poisson brackets (PB).

We can now look at linear functions on $\Gamma$. Denote by $Y^a$
an element of $\Gamma$, and using the fact that it is linear
space, $Y^a$ also represents a vector in $T\Gamma$. Given a one
form $\lambda_a$, we can define a linear function on $\Gamma$ as
follows: $F_\lambda(Y):=-\lambda_aY^a$. Note that $\lambda$ is a
label of the function with $Y^a$ as its argument. First, note that
there is a vector associated to $\lambda_a$:
\[
\lambda^a:=\Omega^{ab}\lambda_b \, ,
\]
so we can write \be
F_\lambda(Y)=\Omega_{ab}\lambda^aY^b=\Omega(\lambda,Y) \, .\ee If
we are now given another label $\nu$, such that
$G_\nu(Y)=-\nu_aY^a$, we can compute the PB \be
\{F_\lambda,G_\nu\}=\Omega^{ab}\nabla_aF_\lambda(Y)\nabla_bG_\nu(Y)\Omega^{ab}\lambda_a\nu_b
\, . \ee Since the two-form is non-degenerate we can rewrite it as
$\{F_\lambda,G_\nu\}=-\Omega_{ab}\lambda^a\nu^b$. Thus, \be
\{\Omega(\lambda,Y),\Omega(\nu,Y)\}=-\Omega(\lambda,\nu) \, .\ee
As we shall see in Sec.~\ref{sec:2.b} we can also make such a
selection of linear functions for the
 Klein-Gordon field.

The quantum representation is the ordinary Schr\"odinger
representation where the Hilbert space is $\H=L^2(R^3,\d^3 x)$ and
the operators are represented by: \be
(\hat{1}\cdot\Psi)(q)=\Psi(q)\qquad(\hat{q^i}\cdot\Psi)(q)=q^i\Psi(q)
\qquad(\hat{p}_i\cdot\Psi)(q)=-i\hbar\frac{\partial}{\partial
q^i}\Psi(q)\, . \label{ordrep}
 \ee

Thus, we recover the conventional quantum theory in the
Schr\"odinger representation.

\subsection{Phase Space and Observables for a scalar field}
\label{sec:2.b}

In this part we shall recall the phase space and Hamiltonian
description of a real, linear Klein-Gordon field  $\phi(x^\mu)$.
 In this
paper we shall assume that we are in Minkowski spacetime, so the
theory we are considering is defined on ${}^4M$. We will perform a
$3+1$ decomposition of the spacetime in the form $M=\Sigma \times
R$, for $\Sigma$ any Cauchy surface, which in this case is
topologically $R^3$. We will consider arbitrary embeddings of the
surfaces $\Sigma$ into   ${}^4M$. The first step is to write the
classical action for the field, \be S:= -{\frac{1}{2}}\int_M
(g^{ab}\nabla_{a}\phi \nabla_{b}\phi
+m^2\phi^2)\;\sqrt{|g|}\,\d^4\! x\, . \label{action} \ee The field
equation is then, \be (\nabla^a\nabla_a - m^2)\,\phi=0
\label{kgeq} \, .\ee Next, we decompose the spacetime metric as
follows: $g^{ab}=h^{ab}-n^an^b$. Here $h^{ab}$ is the (inverse of)
the induced metric on the Cauchy hypersurface $\Sigma$ and $n^a$
the unit normal to $\Sigma$. We also introduce an everywhere
time-like vector field $t^a$ and a `time' function $t$ such that
the hypersurfaces $t=$constant are diffeomorphic to $\Sigma$ and
such that $t^a\nabla_at=1$. Note that, for each $t$, we have an
embedding of the form $T_t:\Sigma\rightarrow {}^4M$. Thus, a
choice of function $t$ provides a one-parameter family of
embeddings (a foliation of ${}^4M$). We can write $t^a=Nn^a+N^a$.
The volume element is given by $\sqrt{|g|}\,\d^4\!
x=N\,\sqrt{h}\,\d^3\! x$.
 Using these identities in Eq.(\ref{action}) we get, \be
S={\frac{1}{2}}\int_I\d t\int_\Sigma
N\sqrt{h}\,\left[(n^a\nabla_a\phi)^2
-h^{ab}\nabla_a\phi\nabla_b\phi-m^2\phi^2\right] \, \d^3\!
x,\label{ac3+1} \ee where $I=[t_0,t_1]$ is an interval in the real
line. Using the relation
\[
n^a\nabla_a\phi=\frac{1}{N}(t^a-N^a)\nabla_a\phi=\frac{1}{N}\dot{\phi}-
\frac{1}{N}N^a\nabla_a\phi \, ,
\]
we can conclude then that the momentum density $\pi$, canonically
conjugate to the configuration variable $\phi$ on $\Sigma$, is
given by \be \pi=\frac{\delta
S}{\delta\dot{\phi}}\sqrt{h}\,(n^a\nabla_a\phi). \ee The phase
space $\G$ of the theory can thus be written as
$\Gamma=(\varphi,\pi)$, where the configuration variable $\varphi$
is the restriction of $\phi$ to $\Sigma$ and $\pi$ is
$\sqrt{h}\,n^a\nabla_a \phi$ restricted to $\Sigma$. Note that the
phase space is of the form $\Gamma=T^*{\cal C}$, where the
classical configuration space ${\cal C}$ can be taken as suitable
initial data (for instance, smooth functions of compact support).

There is an alternative description for the phase space of the
theory, given by the ``covariant phase space''\cite{covariant}. In
this approach, the phase space is the space of {\it solutions}
$\phi$ to the equation of motion. Let us denote this space by
$\Gamma_{\rm s}$. Note that, for each embedding $T_t:\Sigma
\rightarrow {}^4M$, there exists an isomorphism ${\cal I}_t$
between $\Gamma$ and $\Gamma_{\rm s}$. The key observation is that
there is a one to one correspondence between a pair of initial
data of compact support on $\Sigma$, and solutions to the
Klein-Gordon equation on ${}^4M$\cite{wald2}.

Therefore, to each element $\phi$ in $\G_s$ there is a pair
$(\varphi,\pi)$ on $\G$ ($\varphi = T_t^*[\phi]$ and
$\pi=T_t^*[\sqrt{h}n^a\nabla_{a} \phi]$) and conversely, for each
pair, there is a solution to the Klein-Gordon equation that
induces the given initial data on $\Sigma$.

In the phase space $\G$ the symplectic structure $\Omega$ takes
the following form, when acting on vectors $(\varphi_1,\pi_1)$ and
$(\varphi_2,\pi_2)$, \be
 \Omega([\varphi_1,\pi_1],[\varphi_2,\pi_2])=\int_\Sigma(\pi_1\varphi_2-
 \pi_2\varphi_1)\,\d^3x
 \label{symp2} \, .
 \ee

Observables for the space $\G$ can be constructed directly by
giving smearing functions on $\Sigma$. We can define linear
functions on $\G$ as follows: given a vector $Y^\alpha$ in $\G$ of
the form $Y^\alpha=(\varphi,\pi)^\alpha$, and a pair
$\lambda_\alpha=(-f,-g)_\alpha$, where $f$ is a scalar density and
$g$ a scalar, we define the action of $\lambda$ on $Y$ as, \be
F_\lambda(Y)=-\lambda_\alpha
Y^\alpha:=\int_\Sigma(f\varphi+g\pi)\,\d^3x\label{observ} \, .\ee
 Now,
we can write this linear function in the form
$F_\lambda(Y)=\Omega_{\alpha\beta}\lambda^\alpha
Y^\beta=\Omega(\lambda,Y)$, if we identify
$\lambda^\beta=\Omega^{\beta\alpha}\lambda_\alpha=(-g,f)^\beta$.
That is, the smearing functions $f$ and $g$ that appear in the
definition of the observables $F$ and are therefore naturally
viewed as a 1-form on phase space, can also be seen as the vector
$(-g,f)^\beta$. Note that the role of the smearing functions is
interchanged in the passing from a 1-form to a vector. Of
particular importance for what follows is to consider {\it
configuration} and {\it momentum} observables. They are particular
cases of the observables $F$ depending of specific choices for the
label $\lambda$. Let us consider the ``label vector''
 $\lambda^\alpha=(0,f)^\alpha$, which would be normally regarded
as a vector in the ``momentum'' direction. However, when we
consider the linear observable that this vector generates, we get,
\begin{equation}
\varphi[f]:=\int_\Sigma \d^3\! x\,f\,\varphi\, .\label{observ2}
\end{equation}
Similarly, given the vector  $(-g,0)^\alpha$ we can construct,
\begin{equation}
\pi[g]:=\int_\Sigma\d^3\! x\,g\,\pi .\label{observ3}
\end{equation}

Note that any pair of test fields $(-g,f)^\alpha\in \G$ defines a
linear observable, but they are `mixed'. More precisely, a scalar
$g$ in $\Sigma$, that is, a pair $(-g,0)\in \G$ gives rise to a
{\it momentum} observable $\pi[g]$ and, conversely, a scalar
density $f$, which gives rise to a vector $(0,f) \in \G$ defines a
{\it configuration} observable $\varphi[f]$. In order to avoid
possible confusions, we shall make the distinction between {\it
label vectors} $(-g,f)^\alpha$ and {\it coordinate vectors}
$(\varphi,\pi)^\alpha$.

As we have seen, the phase space can be alternatively described by
solutions to the Klein-Gordon equation in the covariant formalism
($\G_{\rm s}$) or by pairs of fields on a Cauchy surface $\Sigma$
in the canonical approach ($\G$). In both cases, the elements of
the algebra $\S$ to be quantized are linear functionals of the
basic fields. In the following sections we consider the
construction of the quantum theory, both in the Fock and in the
Schr\"odinger representation.

\section{Fock Quantization}
\label{sec:3}

Let us now consider  the Fock quantization. The intuitive idea is
that the Hilbert space of the theory is constructed from
``n-particle states". (In certain cases one is justified to
interpret the quantum states as consisting of n-particle states.)
The Fock quantization is naturally constructed from {\it solutions
to the classical equations of motion} and relies heavily on the
linear structure of the space of solutions (The Klein-Gordon
equation is linear). Thus, it can only be implemented for
quantizing {\it linear (free) field theories}. The main steps of
the quantization of the Klein-Gordon field are the following:
Given a 4-dimensional globally hyperbolic spacetime $(M,g)$, the
first step is to consider the vector space $\Gamma_{\rm s}$ of
solutions of the equation of motion. One then constructs the
algebra  $\S$ of fundamental observables to be quantized, which in
this case consists of suitable linear functionals on $\Gamma_{\rm
s}$. The next step is to construct the so called {\it one-particle
Hilbert space} ${\cal H}_0$ from the space $\Gamma_{\rm s}$, as we
will see below the only input that we require to construct ${\cal
H}_0$ is to introduce a complex structure compatible with the
naturally defined symplectic structure on $\Gamma_{\rm s}$.
 As mentioned before, the one-particle Hilbert space ${\cal H}_0$
receives this name since it can be interpreted as the Hilbert
space of a one-particle relativistic system. From the Hilbert
space ${\cal H}_0$ one constructs its symmetric (since we are
considering Bose fields) Fock space ${\cal F}_{s}({\cal H}_0)$,
the Hilbert space of the theory.
The final step is to represent the quantum version of the algebra
$\S$ of observables in the Fock space as suitable combinations of
(naturally defined) creation and annihilation operators.

The classical observables to be quantized are the fundamental
fields which correspond to exact solutions of Eq.(\ref{kgeq}). The
next step in the quantization program is to identify the
one-particle Hilbert space ${\H}_0$. The strategy is the
following: start with $(\G_{\rm s},\w)$ a symplectic vector space
and define $J:\G_{\rm s}\rightarrow\G_{\rm s}$, a linear operator
such that $J^2=-1$. The {\it complex structure} $J$ has to be
compatible with the symplectic structure. This means that the
bilinear mapping defined by $\mu(\cdot,\cdot):=\w(J\cdot,\cdot)$
is a positive definite metric on $\G_{\rm s}$. The Hermitian
(complex) inner product is then given by, \be
\langle\cdot,\cdot\rangle={{1}\over{2\hbar}}\mu(\cdot,\cdot)-
i{{1}\over{2\hbar}}\w(\cdot,\cdot)\, .\label{innp}
 \ee
 The complex structure $J$ defines a natural splitting of $\G_\C$,
the complexification of $\G_{\rm s}$, in the following way: Define
the {\it positive frequency} part to consist of vectors of the
form $\Phi^+:={{1}\over{2}}(\Phi-iJ\Phi)$ and the {\it negative
frequency} part as $\Phi^-:={{1}\over{2}}(\Phi+iJ\Phi)$. Note that
$\Phi^-=\overline{\Phi}^+$ and $\Phi=\Phi^++\Phi^-$. Since
$J^2=-1$, the eigenvalues of $J$ are $\pm i$, so one is
decomposing the vector space $\G_\C$ in eigenspaces of $J$:
$J(\Phi^{\pm})=\pm i\Phi^{\pm}$. We have used the term
`positive-negative frequency' since in the case of the Minkowski
spacetime the naturally defined $J$ induces this standard
decomposition (See Sec.~\ref{sec:6}).

There are two alternative but completely equivalent descriptions
of the one-particle Hilbert space ${\H}_0$:
\begin{itemize}

\item ${\H}_0$ consists of {\it real} valued functions (solution to the
Klein-Gordon equation for instance), equipped with the complex
structure $J$. The inner product is given by (\ref{innp}).

\item ${\H}_0$ is constructed by complexifying the vector space $\G_{\rm s}$
(tensoring with the complex numbers) and then decomposing it using
$J$ as described above. In this construction, the inner product is
given by,
$$
\langle\Phi,\tilde{\Phi}\rangle=-{{i}\over{\hbar}}\w(\Phi^-,\tilde{\Phi}^+)
\, .
$$
Note that in this case, the one-particle Hilbert space consists of
`positive frequency' solutions.

\end{itemize}

It is important to note that the only input we needed in order to
construct ${\H}_0$ was the complex structure $J$.

The {\it symmetric Fock space} associated to ${\H}_0$ is defined
to be the Hilbert space
$$
\F_{\rm s}({\H}_0):=\bigoplus^{\infty}_{n=0}
\left(\bigotimes^n{}_{\rm s}{\H}_0\right)\, ,
$$
where we define the {\it symmetrized tensor product} of ${\H}_0$,
denoted by $\bigotimes^n{}_{\rm s}{\H}_0$, to be the subspace of
the n-fold tensor product ($\bigotimes^n{\H}_0$), consisting of
totally symmetric maps
$\alpha:\overline{\H}_1\times\cdots\times\overline{\H}_n\rightarrow
\C$
 (with ${\H}_1=...={\H}_n={\H}_0=:{\H}$) satisfying
$$\sum \left|\alpha(\bar{e}_{1i_1},\ldots,\bar{e}_{ni_n})\right|^2<\infty\, .
$$
The Hilbert space $\overline{\H}$ is the {\it complex conjugate}
of $\H$ with $\{\bar{e}_{1},\cdots,\bar{e}_{j},\cdots\}$ an
orthonormal basis. We are also defining $\bigotimes^0\H=\C$.

We shall introduce the abstract index notation for the Hilbert
spaces since it is the most convenient way of describing the Fock
space. Given a space $\H$, we can construct the spaces
$\overline{\H}$, the complex conjugate space; $\H^*$, the {\it
dual space}; and $\overline{\H}^*$ the dual to the complex
conjugate. In analogy with the notation used in spinors, let us
denote elements of $\H$ by $\phi^A$, elements of $\overline{\H}$
by $\phi^{A^\prime}$. Similarly, elements of ${\H}^*$ are denoted
by $\phi_A$ and elements of $\overline{\H}^*$ by
$\phi_{A^\prime}$. However, by using Riesz lemma, we may identify
$\overline{\H}$ with $\H^*$ and $\H$ with $\overline{\H}^*$.
Therefore we can eliminate the use of primed indices, so
$\overline{\phi}_{A}$ will be used for an element in
$\overline{\H}^*$ corresponding to the element $\phi^A\in \H$. An
element $\phi\in\bigotimes^n{}_{\rm s}\H$ then consists of
elements satisfying \be \phi^{A_1\cdots A_n}=\phi^{(A_1\cdots
A_n)}. \ee An element $\psi\in \bigotimes^n\overline{\H}$ will be
denoted as $\psi_{A_1\cdots A_n}$. In particular, the inner
product of vectors $\psi,\phi\in \H$ is denoted by
\[
\langle\psi,\phi\rangle=:\overline{\psi}_A\phi^A \, .
\]

A vector $\Psi\in\F_{\rm s}(\H)$ can be represented, in the
abstract index notation as
\[
\Psi=(\psi,\psi^{A_1},\psi^{A_1A_2},\ldots ,\psi^{A_1\ldots A_n},
\ldots )\, ,
\]
where, for all $n$, we have $\psi^{A_1\ldots A_n}=\psi^{(A_1\ldots
A_n)}$. The norm is given by \be
|\Psi|^2:=\overline{\psi}\psi+\overline{\psi}_A\psi^A+\overline{\psi}_{A_1A_2}
\psi^{A_1A_2}+\cdots < \infty\, . \ee
 Now, let $\xi^A\in\H$ and
let $\overline{\xi}_A$ denote the corresponding element in
$\overline{\H}$. The {\it annihilation operator} ${\cal
A}(\bar{\xi}): \F_{\rm s}(\H)\rightarrow \F_{\rm s}(\H)$
associated with $\overline{\xi}_A$ is defined by \be {\cal
A}(\bar{\xi})\cdot\Psi:=(\overline{\xi}_A\psi^A,\sqrt{2}\,
\overline{\xi}_A
\psi^{AA_1},\sqrt{3}\,\overline{\xi}_A\psi^{AA_1A_2},\ldots )\, .
\label{aniqui}
 \ee
 Similarly, the {\it creation operator} ${\cal C}({\xi}):
\F_{\rm s}(\H)\rightarrow \F_{\rm s}(\H)$ associated with $\xi^A$
is defined by \be {\cal
C}(\xi)\cdot\Psi:=(0,\psi\xi^{A_1},\sqrt{2}\,\xi^{(A_1}\psi^{A_2)},
\sqrt{3}\,\xi^{(A_1}\psi^{A_2A_3)},\ldots )\, .\label{crea}
 \ee
 If the domains of the operators are defined to be the
subspaces of $\F_{\rm s}(\H)$ such that the norms of the right
sides of Eqs. (\ref{aniqui}) and (\ref{crea}) are finite then it
can be proven that ${\cal C}(\xi)=({\cal A}(\bar{\xi}))^\dagger$.
It may also be verified that they satisfy the commutation
relations, \be \left[{\cal A}(\bar{\xi}),{\cal
C}(\eta)\right]=\bar{\xi}_A\eta^A\,{\rm \hat{I}}\, . \label{CCR2}
\ee A more detailed treatment of Fock spaces can be found in
\cite{wald2,wald1,geroch,reeds}.

In the previous section we saw that we could construct linear
observables in $(\G,\w)$, which we denoted by $F_\lambda(Y)$.
These observables are given by (\ref{observ}) and in the more
canonical picture by (\ref{observ2}) and (\ref{observ3}). This is
the set ${\cal S}$ of observables, now denoted by ${\cal
O}[\eta]$, for which there will correspond a quantum operator.
Thus, for ${\cal O}[\eta]\in {\cal S}$ there is an operator
$\hat{\cal O}[\eta]$. We want the CCR to hold,
\begin{equation}
\left[\hat{\cal O}[\eta],\hat{\cal O}[\xi]\right]=i\hbar \{{\cal
O}[\eta],{\cal O}[\xi]\}\, {\rm \hat{I}}=-i\hbar\,\w(\eta,\xi)\,
{\rm \hat{I}}\, .
\end{equation}
Then we should see that the Fock construction is a Hilbert space
representation of our basic operators satisfying the above
conditions. We have all the structure needed at our disposal. Let
us take as the Hilbert space the symmetric Fock space $\F_{\rm
s}(\H)$ and let the operators be represented as
\begin{equation}
\hat{\cal O}[\eta]\cdot\Psi:=i\hbar\left({\cal A}(\overline{\eta})
- {\cal C}(\eta) \right) \cdot \Psi\, .\label{repsobs}
\end{equation}
Let us denote by $\eta^A$ the abstract index representation
corresponding to the pair $(-g,f)$ in $\H$. That is, the vector
$\eta$ appearing in the previous expressions should be thought of
as a {\it label} for the state created (or annihilated) by ${\cal
C} (\eta)$ (${\cal A}(\overline{\eta})$) on the Hilbert space.

Let us now focus on the properties of the operators given by
(\ref{repsobs}). First, note that by construction the operator is
self-adjoint. It is straightforward to check that the commutation
relations are satisfied, \ba \left[\hat{\cal O}[\eta],\hat{\cal
O}[\xi]\right]&=& \hbar^2\big[{\cal A}(\overline{\eta}), {\cal
C}(\xi)\big]+ \hbar^2\big[{\cal C}(\eta),{\cal
A}(\overline{\xi})\big]\nl &=&\hbar^2\,(
\overline{\eta}_A\xi^{A}-\overline{\xi}_A\eta^A)\, {\rm \hat{I}}
\nl &=&\hbar^2\,(\langle \eta, \xi\rangle-\langle
\xi,\eta\rangle)\, {\rm \hat{I}}\nl &=& 2i\hbar^2\,{\rm
Im}(\langle \eta,\xi\rangle) \, {\rm
\hat{I}}=-i\hbar\,\w(\eta,\xi) \, {\rm \hat{I}}\, , \ea where we
have used (\ref{CCR2}) in the second line
 and (\ref{innp}) in the last line. Note that in this last calculation
we only used general properties of the Hermitian inner product and
therefore we would get a representation of the CCR for {\it any}
inner product $\langle\cdot,\cdot\rangle$. Since the inner product
is given in turn by a complex structure $J$, we see that there is
a one to one correspondence between them, and that $J$ represents
the (infinite) freedom in the choice of the quantum
representation.

\section{Schr\"odinger Representation}
\label{sec:4}

 In the previous section we considered the Fock
quantization of the Klein-Gordon field,  one of its most notable
features being the fact that it is most naturally stated and
constructed in a covariant framework. In particular, the
symplectic structure, even when it uses explicitly a hypersurface
$\Sigma$, is independent of this choice. The same is true for the
complex structure which is a mapping from solutions to solutions.
The infinite dimensional freedom in choice of representation of
the CCR relies in the choice of admissible $J$, which gives rise
to the one-particle Hilbert space. Thereafter, the construction is
completely natural and there are no further choices to be made. We
know that from the infinite possible choices of $J$ there are
physically inequivalent representations \cite{wald2}, a clear
indication that the Stone-von Neumann theorem does not generalize
to field theories.

We now turn our attention to the Schr\"odinger representation. In
contrast to the previous case, this construct relies heavily on a
Cauchy surface $\Sigma$, since its most naive interpretation is in
terms of a ``wave functional at time $t$". For simplicity, we have
assumed that we are in Minkowski spacetime, so the theory we are
considering is defined on ${}^4M$ in the form $M=\Sigma \times R$.
However, we are considering arbitrary embeddings, so the surface
$\Sigma$ is topologically $R^3$, but can have an arbitrary metric
$h_{ab}$, and extrinsic curvature on it. Recall that the phase
space of the theory can be written as $\Gamma=(\varphi,\pi)$,
where $\varphi = T_t^*[\phi]$ and $\pi=T_t^*[\sqrt{h}n^a\nabla_{a}
\phi]$.
 Recall that in this case the phase space is of the form
$\Gamma=T^*{\cal C}$, where the classical configuration space
${\cal C}$ can be taken as suitable initial data (for instance,
smooth functions of compact support). The results reported in this
section follow \cite{nosotros} closely.

\subsection{First Steps}
\label{sec:4.a}

 The Schr\"odinger representation, at least in an
intuitive level, is to consider `wave functions' as function(al)s
of $\varphi$. That is \be {\cal H}_{\rm s}:=L^2(\overline{\cal C},
\d\mu) \label{hspace} \, ,\ee where a state would be represented
by a function(al) $\Psi[\varphi]: \overline{\cal C}\rightarrow
\C$.

We have already encountered two new actors in the play. First
comes the {\it quantum configuration space} $\overline{\cal C}$,
and the second one is the measure $\mu$ thereon. Thus, one will
need to specify these objects in the construction of the theory.
Before going into that, let us look at the classical observables
that are to be quantized, and in terms of which the CCR are
expressed.
 Recall the observables (\ref{observ2}) and (\ref{observ3}),
\[
\varphi[f]:=\int_{\Sigma} \d^3\!x\;f\,\varphi
\]
and
\[
\pi[g]:=\int_{\Sigma} \d^3\!x\;g\,\pi \, ,
\]
where the test functions $g$ and $f$ are again smooth and of
compact support. The Poisson bracket between the configuration and
momentum observables is, \be \{\varphi[f],\pi[g]\}=\int
\d^3\!x\;f\,g\, , \ee
 so the canonical commutation relations read
$[\hat{\varphi}[f],\hat{\pi}[g]] =i\hbar\int \d^3\!x\; fg \, {\rm
\hat{I}}$.

Recall that in the general quantization procedure the next step is
to represent the abstract operators $\hat{\varphi}[f]$ and
$\hat{\pi}[g]$ as operators in a Hilbert space, with the
appropriate ``reality conditions", which in our case means that
these operators should be Hermitian.

We can represent them, when acting on functionals $\Psi[\varphi]$
as \be
(\hat{\varphi}[f]\cdot\Psi)[\varphi]:=\varphi[f]\,\Psi[\varphi]\,
,\label{comfop}
 \ee
 and
 \be (\hat{\pi}[g]\cdot\Psi)[\varphi]:=-i\hbar\int
\d^3\!x\;g(x) {{\delta \Psi}\over{\delta \varphi(x)}} + {\rm
multiplicative\; term}\, .\label{momop}
 \ee
 The second term in (\ref{momop}), depending only on
configuration variable is there to render the operator
self-adjoint when the measure is different from the ``homogeneous"
measure, and depends on the details of the measure. At this point
we must leave it unspecified since we have not defined the measure
yet.

In quantum mechanics, we are used to the fact that we can simply
take the same measure (Lebesgue) on $R^3$ for an ordinary problem
and not even worry about this issue. We are saved in that case by
the Stone-von Neumann theorem that assures us that any `decent'
representation of the CCR is unitarily equivalent to the
Schr\"odinger one. In field theory this is false. There are
infinitely inequivalent representations of the CCR. In the Fock
representation we saw that the ambiguity is encoded in the complex
structure $J$. However, in the Schr\"odinger picture we encounter
the first conceptual difficulty. How does the infinite ambiguity
existent manifest itself in the Schr\"odinger picture?
Intuitively, one expects that the information be somehow encoded
in the measure $\mu$, which will, by the reality conditions,
manifest itself in the choice of the representation of the
momentum operator (\ref{momop}). This intuitive picture gets
entangled however with two features that were absent in the Fock
construction. The first one is that one is tempted to apply an old
trick which is very useful in, say, a harmonic oscillator in QM.
Recall that in that case, one can either consider the Hilbert
space $L^2(R,\d q)$ of square integrable functions with respect to
the Lebesgue measure $\d q$, in which case one has the
representation of the operators as in (\ref{ordrep}). The vacuum
is given by a Gaussian of the form $\psi_0(q)=\exp(-q^2/2)$. The
other possibility is to ``incorporate the vacuum" into the
measure, in such a way that the measure becomes now
$\d\mu^\prime=\exp(-q^2)\d q$, and the vacuum is the unit function
$\tilde{\psi}_0(q)=1$. With this choice, the momentum operator
acquires an extra term such that $\hat{\tilde{p}}=-i\hbar(\d/\d
q-\frac{1}{2}\,q)$. The two representations are completely
equivalent since one can go from any wave-function $\psi(q)$ in
the standard representation to a state in the new representation
by $\psi \mapsto \tilde{\psi}(q)=\exp(q^2/2)\psi$. This map is a
unitary isomorphism of Hilbert spaces and thus the two
representations are equivalent. The question now is whether one
can apply such a map in the functional case, and go from a
`simple' representation with uniform measure to a `complicated'
representation with a non-uniform measure, in a unitary way. (See
\ref{app:b} for the construction of the vacuum state for the case
of the Klein-Gordon field.)

The second feature that was briefly mentioned before has to do
with the fact that one is defining the theory on a particular
Cauchy surface and therefore there might be extra complications
coming from a non-standard choices of embeddings. That is, in the
covariant picture the complex structure is independent of any such
hypersurface, and might induce very different looking maps, in
terms of initial data, for different choices of $\Sigma_t$. In
what follows we shall see that one can overcome these difficulties
and define a {\it canonical} representation of the CCR, in terms
of what is normally referred to as the associated {\it Gaussian
measure}.

In order to understand the situation, let us explore what appears
to be the simplest and most attractive possibility, namely let us
consider the case in which the measure is the uniform one. That
is, the measure can be written something like ``$\d\mu={\cal
D}\varphi$", and would be the equivalent of the Lebesgue measure
in the real line. In this case, the momentum operator is
represented as follows: \be \hat{\pi}[g]=-i\hbar\int \d^3\!x\;
g(x){{\delta}\over{\delta \varphi(x)}} \, .\ee From our experience
with the harmonic oscillator, we expect that the vacuum be a
Gaussian wave function (see \ref{app:b}). If we regard the
quantization as a recipe for producing a Hilbert space and a
representation of the basic operators, we are finished! There is
however something very puzzling about this construct. It appears
to be universal for any scalar field theory over $\Sigma$; there
is no trace of the ambiguity present in the Fock picture, namely
of the complex structure $J$. Have we somehow been able to get
something as a ``free lunch", and been able to circumvent the
difficulty? The answer is in the negative and there are at least
two aspects to it. Heuristically, we can see that we have not been
completely successful in the construction, since we have failed to
provide the vacuum state $\Psi_0(\varphi)$. This suggests that the
information about the $J$, and therefore, the different possible
``representations" is encoded in the choice of vacuum and {\it
not} in the representation itself. If this were the case, then
inequivalence of the different representations would manifest
itself as the impossibility to define a unitary map connecting the
different vacuum states. This explanation seems plausible and, as
we shall see later, has some concrete use. In fact this is the
standard issue of ``choice of the vacuum" that one finds in
standard texts \cite{birrell}. However, there is still a deeper
reason why this naive representation is `wrong'. From a technical
point of view, this representation with a ``uniform" measure is
not well defined for the simple reason that such measure does not
exist!

The theory of measures on infinite dimensional vector spaces (such
as the space of initial conditions) has some subtleties, among
which is the fact that well-defined measures should be {\it
probability} measures (this means that $\int_{V}\d\mu=1$)
\cite{draft}. A uniform measure would not have such property. It
is convenient to digress a bit and introduce some basic concepts
from measure theory. In the case of infinite dimensional vector
spaces $V$, there is an object, called the {\it Fourier Transform}
of the measure $\mu$. It is defined as
\[
\chi_\mu(f):=\int_V\d\mu\,e^{if(\varphi)} \, ,
\]
where $f(\varphi)$ is an arbitrary continuous function(al) on $V$.
It turns out that under certain technical conditions, the Fourier
transform $\chi$ characterizes completely the measure $\mu$. This
fact is particularly useful for us since it allows to give a
precise definition of a Gaussian measure. Let us assume that $V$
is a Hilbert space and $B$ a positive-definite, self-adjoint
operator on $V$. Then a measure $\mu$ is said to be Gaussian if
its Fourier transform has the form, \be
\chi_\mu(f)=\exp\left(-\frac{1}{2}\la f,Bf\ra_V\right)\,
,\label{gaussian}
 \ee
where $\la\cdot,\cdot\ra_V$ is the Hermitian inner product on $V$.
We can, of course, ask what the measure $\mu$ looks like. The
answer is that, schematically it has the form, \be ``\d\mu =
\exp\left(-\frac{1}{2}\la\varphi,B^{-1}\varphi\ra_V\right){\cal
D}\varphi"\, ,\label{false}
 \ee
  where ${\cal D}\varphi$ represents
the fictitious ``Lebesgue-like" measure on $V$. The expression
(\ref{false}) should be taken with a grain of salt since it is not
completely well defined (whereas (\ref{gaussian}) is). It is
nevertheless useful for understanding where the denomination of
Gaussian comes from. This can be seen from the term
$-\frac{1}{2}\la\varphi,B^{-1}\varphi\ra_V$ that is (finite and)
negative definite, thus endowing $\mu$ with its Gaussian
character.

Let us return to the previous discussion regarding the
representation of the CCR. We have argued that the trivial
representation, given by a ``uniform measure'' is non-existent,
and furthermore one is forced to consider a probabilistic measure.
Notice that other than being a probabilistic measure, we have no
further restrictions on what the measure $\mu$ should be. It is a
part of ``folklore", in the theoretical physics community, that
the correct measure for our case is Gaussian. It is precisely one
of the purposes of this article to motivate and illustrate this
widely known result. Therefore, we shall try to take the most
straight logical path to the desired result. What we need to do is
to find the measure $\mu_F$ that corresponds to the Fock
representation. That is, given a Fock Hilbert space ${\cal H}_F$,
we want to find the Schr\"odinger Hilbert space that is
``equivalent" to it. So far, we have not been precise about what
we mean by being equivalent. Once we use the proper setting for
specifying ``equivalence" of Hilbert spaces the right measure will
be straightforward to find.

Let us summarize our situation. We saw that in order to construct
the Fock quantum theory, in addition to the naturally defined
symplectic structure on phase space, we needed to specify an
additional {\it classical} structure, namely a complex structure
$J$ on phase space. Furthermore, we have concluded that we need to
specify a measure on the function space $V$ for the Schr\"odinger
representation. The natural strategy is then to try to use the
information that the complex structure $J$ provides, and employ it
for finding the correct measure. We will see in the next section
that $J$ provides us with precisely the right structure needed for
the quantum equivalence notion that the Algebraic Formulation of
Quantum Field Theory defines.

\subsection{Algebras and States}
\label{sec:4.b}

 The question we want to address is how to
formulate equivalence between the two representations, namely Fock
and Schr\"odinger for the scalar field theory. The most natural
way to define this notion is through the algebraic formulation of
QFT (see \cite{haag} and \cite{wald2} for introductions). The main
idea is to formulate the quantum theory in such a way that the
observables become the relevant objects and the quantum states are
``secondary". Now, the states are taken to ``act" on operators to
produce numbers. For concreteness, let us recall the basic
constructions needed.

The main ingredients in the algebraic formulation are two, namely:
(1) a $C^*$-algebra ${\cal A}$ of observables, and (2) states
$\omega:{\cal A}\rightarrow \C$, which are positive linear
functionals ($\omega(A^*A)\geq0\,\forall A\in{\cal A}$) such that
$\omega(1)=1$. The value of the state $\omega$ acting on the
observable $A$ can be interpreted as the expectation value of the
operator $A$ on the state $\omega$, i.e. $\la A\ra=\omega(A)$.

For the case of a linear theory, the algebra one considers is the
so-called {\it Weyl algebra}. Each generator $W(\lambda)$ of the
Weyl algebra is the ``exponentiated" version of the linear
observables (\ref{observ}), labeled by a phase space vector
$\lambda^a$. These generators satisfy the Weyl relations: \be
W(\lambda)^{*}=W(-\lambda)\: , \:\:\:\:
W(\lambda_{1})W(\lambda_{2})=e^{{{i}\over{2}}\Omega(\lambda_{1},\lambda_{2})}
W(\lambda_{1}+ \lambda_{2}) \, .\ee

The CCR get now replaced by the quantum Weyl relations where now
the operators $\hat{W}(\lambda)$ belong to the (abstract) algebra
${\cal A}$. Quantization in the old sense means a representation
of the Weyl relations on a Hilbert space. The relation between
these concepts and the algebraic construct is given through the
GNS construction that can be stated as the following theorem
\cite{wald2}:

{\it{Let ${\cal A}$ be a $C^*$-algebra with unit and let
$\omega:{\cal A}\rightarrow \C$  be a state. Then there exist a
Hilbert space ${\cal H}$, a representation $\pi:{\cal
A}\rightarrow L({\cal H})$ and a vector $|\Psi_0\ra\in {\cal H}$
such that, \be \omega(A)=\la \Psi_0,\pi(A)\Psi_0\ra_{\cal H}
\label{teo-expect} \, .\ee Furthermore, the vector $|\Psi_0\ra$ is
cyclic. The triplet $({\cal H},\pi,|\Psi_0\ra)$ with these
properties is unique (up to unitary equivalence)}}.

One key aspect of this theorem is that one may have different, but
unitarily equivalent, representations of the Weyl algebra, which
will yield {\it equivalent} quantum theories. This is the precise
sense in which the Fock and Schr\"odinger representations are
related to each other. Let us be more specific. We have in
previous sections constructed a Fock representation of the CCR,
with the specification of a complex structure $J$. Using this
representation, we can now compute the expectation value of the
Weyl operators on the Fock vacuum and thus obtain a positive
linear functional $\omega_{\rm fock}$ on the algebra ${\cal A}$.
Now, the Schr\"odinger representation that will be equivalent to
the Fock construction will be the one that the GNS construction
provides for the {\it same} algebraic state $\omega_{\rm fock}$.
Our job now is to complete the Schr\"odinger construction such
that the expectation value of the corresponding Weyl operators
coincide with those of the Fock representation.

The first step in this construction consists in writing the
expectation value of the Weyl operators in the Fock representation
in terms of the complex structure $J$. The action of the state
$\omega_{\rm fock}$ on the Weyl algebra elements
$\hat{W}(\lambda)$ is given by \cite{baez,wald2,nosotros}
 \be
\omega_{\rm
fock}(\hat{W}(\lambda))=e^{-\frac{1}{4}\mu(\lambda,\lambda)}\label{magic}
\, ,
 \ee
where $\mu(\cdot,\cdot):=\Omega(J\cdot,\cdot)$ is the positive
definite metric defined on the phase space.

\subsection{Measure and Representation}
\label{sec:4.c}

The next step is to complete the Schr\"odinger representation,
which is now a two step process. First we need to find the measure
$\d\mu$ on the quantum configuration space in order to get the
Hilbert space (\ref{hspace}) and second we need to find the
representation (\ref{comfop}) and (\ref{momop}) of the basic
operators.

Let us write the complex structure $J$ in terms of the initial
data. On the phase space $(\Gamma, \Omega)$ with coordinates
$(\varphi,\pi)$, the most general form of the complex structure
$J$ is given by
\begin{equation}
\label{est.com} -J_{\Gamma}(\varphi , \pi)=(A \varphi + B \pi ,
C\pi +D\varphi) \, ,
\end{equation}
where $A,B,C$ and $D$ are linear operators  satisfying the
following relations \cite{6}:
\begin{equation}
\label{relaciones1} A^{2}+BD=-{\bf{1}} \:\: , \:\:\:
C^{2}+DB=-{\bf{1}} \:\: , \:\:\: AB+BC=0 \:\: , \:\:\: DA+CD=0 \,
.
\end{equation}
The inner product $\mu_{\Gamma}(\, \cdot \, , \, \cdot \,)=\w(\,
\cdot \, , -J_{\G} \, \cdot \,)$ in terms of these operators is
given by
\begin{equation}
\label{pint-operadores} \mu_{\Gamma}((\varphi_{1} ,
\pi_{1}),(\varphi_{2}, \pi_{2}))= \int_{\Sigma}\d^3x \,(\pi_{1} B
\pi_{2} + \pi_{1} A \varphi_{2} - \varphi_{1} D
\varphi_{2}-\varphi_{1} C \pi_{2} ) \, ,
\end{equation}
for all pairs $(\varphi_{1} , \pi_{1})$ and $(\varphi_{2},
\pi_{2})$ in $\G$. As $\mu_{\Gamma}$ is symmetric and positive
definite, then the linear operators  should also satisfy \cite{6}
\begin{equation}
\label{relaciones2} \int_{\Sigma} fBf' = \int_{\Sigma} f'B f \:\:
, \:\:\: \int_{\Sigma} gDg' = \int_{\Sigma} g'D g  \:\: , \:\:\:
\int_{\Sigma} fAg = -\int_{\Sigma} gCf
\end{equation}
and \be \int_{\Sigma}fBf'>0 \:\: , \:\:\: \int_{\Sigma}gDg'<0 \,
,\ee for all scalars $g,g' \in C^{\infty}_{0}(\Sigma)$  and unit
weight scalars densities $f,f' \in C^{\infty}_{0}(\Sigma)$.

With this in hand, we can find the measure $\d\mu$ that defines
the Hilbert space. In order to do this, it suffices to consider
configuration observables. That is, we shall consider observables
of the form $\varphi[f]= \int\d^3x\,f\,\varphi$, which correspond
to a vector of the form $\lambda^a=(0,f)^a$. Now, we know how to
represent these observables {\it independently} of the measure
since they are represented as multiplication operators as given by
(\ref{comfop}). In the Schr\"{o}dinger picture, the Weyl
observable $\hat{W}(\lambda)$ with label $\lambda^{a}=(0,f)^a$ has
the form \be \hat{W}_{\rm{sch}}(\lambda)=e^{i \hat{\varphi}[f]} \,
.\ee Now, the equation (\ref{magic}) tells us that the state
$\omega_{\rm sch}$ should be such that \be
\omega_{\rm{sch}}(\hat{W}(\lambda))=\exp\left[-\frac{1}{4}
\int_\Sigma\d^3x\,f\,B\,f\right] \label{VEV1} \, , \ee where we
have used the inner product (\ref{pint-operadores}). On the other
hand, the left hand side of (\ref{magic}) is the vacuum
expectation value of the $\hat{W}(\lambda)$ operator. That is, \be
\omega_{\rm{sch}}(\hat{W}(\lambda))=\int_{\overline{\cal C}}\d\mu
\,
\overline{\Psi_0}(\hat{W}_{\rm{sch}}(\lambda)\cdot\Psi_0)=\int_{\overline{\cal
C}}\d\mu\,e^{i\int_\Sigma \d^3x\,f\,\varphi} \, .\label{VEV2} \ee
Let us now compare (\ref{VEV1}) and (\ref{VEV2}), \be
\int_{\overline{\cal C}}\d\mu\,e^{i\int_\Sigma
\d^3x\,f\,\varphi}=\exp\left[-\frac{1}{4}\int_\Sigma\d^3x\,f\,B\,f\right]
\label{gauss2}\, .\ee

Let us now recall our previous discussion regarding the Fourier
transform of a Gaussian measure, given by Eq. (\ref{gaussian}).
Then we note that (\ref{gauss2}) tells us that the measure $\d\mu$
is Gaussian and that it corresponds heuristically to a measure of
the form \be ``\d\mu=e^{-\int_{\Sigma}\varphi
B^{-1}\varphi}\;{\cal D}\varphi" \label{medida} \, .\ee

This is the desired measure. However, we still need to find the
``multiplicative term" in the representation of the momentum
operator (\ref{momop}). For that, we will need the full Weyl
algebra and Eq. (\ref{magic}). Let us write the most general
momentum operator as ($\hbar = 1$),
\begin{equation}
\label{conf-mom-op} (\hat{\pi}[g]\cdot \Psi)[\varphi]=-i\,
\int_{\Sigma} \biggl(g {{\delta}\over{\delta \varphi}} \biggr)
\Psi[\varphi] + \hat{M}\cdot  \Psi[\varphi] \, .
\end{equation}
Imposing (\ref{magic}) and using the Baker-Campbell-Hausdorff
(BCH) relation, it can be shown that the momentum operator is
uniquely given by the expression \cite{nosotros}
\begin{equation}
\label{repdepi-conC} (\hat{\pi}[g]\cdot
\Psi)[\varphi]=-i\int_{\Sigma} \biggl(g {{\delta}\over{\delta
\varphi}} -\varphi(B^{-1}-iCB^{-1})g \biggr) \Psi[\varphi] \, ,
\end{equation}
which explicitly exhibits the form of the multiplicative term.

To summarize, we have used the vacuum expectation value condition
(\ref{magic}) in order to construct the desired Schr\"odinger
representation, namely, a unitarily equivalent representation of
the CCR on the Hilbert space defined by functionals of initial
conditions. We have provided the most general expression for the
quantum Schr\"odinger theory, for arbitrary embedding of $\Sigma$
into ${}^4M$. From the discussion of Sec.~\ref{sec:4.a}, we saw
that the only possible representation was in terms of a
probability measure, thus ruling out the naive ``homogeneous
measure". This conclusion made us realize that both the choice of
measure and the representation of the momentum operator were
intertwined; the information about the complex structure $J$ that
lead to the ``one-particle Hilbert space" had to be encoded in
both of them. We have shown that the most natural way to put this
information as conditions on the Schr\"odinger representation was
through the condition (\ref{magic}) on the vacuum expectation
values of the basic operators. This is the non-trivial input in
the construction.

Several remarks are in order. \begin{enumerate}
 \item  In
Sec.~\ref{sec:4.a} we made the distinction between the classical
configuration space ${\cal C}$ of initial configurations
$\varphi(x)$ of compact support and the quantum configuration
space $\overline{\cal C}$. So far we have not specified
$\overline{\cal C}$. In the case of flat embeddings, where
$\Sigma$ is a Euclidean space, the quantum configuration space is
the space ${\cal J}^*$ of tempered distributions on $\Sigma$.
However, in order to define this space one uses the linear and
Euclidean structure of $\Sigma$ and it is not trivial to
generalize it to general curved manifolds. This subtleties lie
outside the scope of this paper, and will be reported elsewhere
\cite{ccq:qcs}.

\item  Note that the form of the measure given by (\ref{gauss2})
is always Gaussian. This is guarantied by the fact that the
operator $B$ is positive definite in the ordinary $L^2$ norm on
$\Sigma$, whose proof is given in \cite{6}. However, the
particular realization of the operator $B$ will be different for
different embeddings $T_t$ of $\Sigma$. Thus, for a given $J$, the
explicit form of the Schr\"odinger representation depends, of
course, on the choice of embedding.

\item  In the discussion regarding quantization in section
\ref{sec:2}, we saw that the operator $\hat{\pi}[g]$ should be
Hermitian. It is straightforward to show that the operator given
by (\ref{repdepi-conC}) indeed satisfies this requirement.

\item  The last term in
the representation of the momentum operator containing the $C$
operator, is somewhat unexpected; it can not be guessed from the
form of the measure, that knows only about $B$. One might hope
that one can get rid of this term by a suitable canonical
transformation. However, it has been shown that the absence of
this term in the momentum operator (i.e. same $B$ and $C=0$) might
lead to inequivalent quantum theories \cite{ccq:cqg}, and examples
of such cases have been explicitly constructed \cite{jero:phd}.

\item
 Note that the
presence of $B$ and $C$ in the momentum operator
(\ref{repdepi-conC}) emphasize the ambiguity inherent in field
theory. However, for a scalar field propagating on a Minkowski
background there is one preferred complex structure (namely, that
one selected by requiring Poincar\'e invariance) and hence a
preferred Schr\"{o}dinger representation. We shall come back to
this issue in later sections.

\item
We have mentioned in the previous discussion that one can
alternatively interpret the quantization ambiguity in the choice
of different vacua. In the \ref{app:b} these vacua are constructed
in the general case. It is shown that they are closely related to
the measure, but also know about the extra information, namely the
operator $C$.

\end{enumerate}

In previous sections we have successfully answered the question of
finding a Schr\"odinger representation unitarily equivalent to a
given Fock representation. However, the precise relation between
them, i.e. a mapping between states is still missing. That is the
purpose of the following section.



\section{From Fock to Schr\"odinger}
\label{sec:5}

The question we want to address is how the two representations are
related, and how can we pass from one to the other. Let us
consider the simplest case which is the passage from the Fock
representation to the Schr\"odinger. We begin with the preferred,
cyclic state, namely the vacuum $|0\rangle$. Let us denote by
`Kets' the elements of the abstract Hilbert space of states, and
use the brackets for states in some representation. Then $\langle
\varphi|0\rangle$ represents the vacuum in the Schr\"odinger
representation. Let us denote by $\langle \vec{n} |0\rangle$ the
vacuum in the Fock representation, using a notation in analogy to
the n-particle states of, say, a harmonic oscillator.

The first step in this direction is to assume that we have the
Fock states, and both representations. Then, our strategy will be
the following: represent the creation and annihilation operators
acting on wave functionals. If we manage to do this we would be
able to have a way of converting a Fock state into a Schr\"odinger
state. For, almost all states on the Fock space can be generated
by acting, with suitable creation operators on the vacuum. Thus,
by acting on the Schr\"odinger vacuum we would be able to create a
dense subset of states. This strategy assumes that we know what
the vacuum in the Schr\"odinger representation is. From the
discussion in Sec.~\ref{sec:4.a} we know that the Schr\"odinger
vacuum $\langle \varphi|0\rangle$ is given (up to a constant
phase) by the constant function \be \Psi_0(\varphi):=\la \varphi
|0\ra = 1. \ee

The next step is to represent creation and annihilation operators
on ${\cal H}_{\rm s}$. This is given by the following expression.
If we represent by $\zeta^a=(-g,f)\in\Gamma$ a label-vector in the
phase space, we can define the corresponding observable ${\cal
O}_\zeta=\varphi[f]+\pi[g]$ and therefore, a quantum observable
$\hat{{\cal O}}(\zeta)$. We can now recover the creation and
annihilation operators as follows,
 \be
{\cal C}(\zeta):={{1}\over{2\hbar}}(\hat{\cal O}(J\zeta)
+i\hat{\cal O} (\zeta)) \ee and \be {\cal
A}(\zeta):={{1}\over{2\hbar}}(\hat{\cal O}(J\zeta) -i\hat{\cal O}
(\zeta)) \, .
 \ee

Recall that the complex structure $J$ acts on initial data as
$-J(\varphi,\pi)=(A\varphi+B\pi,C\pi+D\varphi)$. Then we have, \be
{\cal C}(-g,f)={{1}\over{2\hbar}}\left(\hat{\varphi}[Dg-Cf+if] +
\hat{\pi}[Bf-Ag+ig]\right)\label{creac} \, .
 \ee
The annihilation operator can be written in a similar way, \be
{\cal A}(-g,f)={{1}\over{2\hbar}}\left(\hat{\varphi}[Dg-Cf-if]+
\hat{\pi}[Bf-Ag-ig]\right)\label{aniquil} \, .
 \ee
These expressions (\ref{creac}) and (\ref{aniquil}) are completely
general, for any $J$ and any representation. In the particular
case we are interested, namely when the representation is
equivalent to the Fock one and is given by (\ref{comfop}) and
(\ref{repdepi-conC}), then we have the desired operators. Note
that in order to have a consistent formulation we should have that
\be {\cal A}(-g,f)\cdot\Psi_0[\varphi]=0,
 \label{vacdef}\ee
for all $f$ and $g$. It is straightforward to check that this is
indeed the case.

We can also find the ``one-particle state" in the Schr\"odinger
representation, which we will denote by $\Phi^1_\zeta:={\cal
C}[\zeta]\cdot\Psi_0$. Using the creation operator (\ref{creac})
on the vacuum state we have that
 \be
\Phi^1_\zeta[\varphi]={{i}\over{\hbar}}\left(\int\d^3\!x\;
\varphi[f+i(B^{-1}-iCB^{-1})g] \right)
 \ee
 is the ``one particle
state" given by the vector $\zeta=(-g,f)$. Furthermore, any state
in the Schr\"odinger representation can be obtained by
successively acting with the creation operator (\ref{creac}).

In the functional picture the two-point function corresponds to
the $L^{2}(\overline{\cal C},\d\mu)$-inner product  between the
``one-particle''
 Schr\"{o}dinger states $\Phi^1_\zeta(\varphi)$ and $\Phi^1_\eta(\varphi)$,
\be \label{2p-function} \la \Phi^1_\zeta(\varphi) ,
\Phi^1_\eta(\varphi) \ra = \bar{\zeta}_A\eta^A
=\frac{1}{2\hbar}\mu (\zeta , \eta) - \frac{i}{2\hbar}\w(\zeta ,
\eta) \, . \ee More precisely, the specification of an algebraic
state $\omega$ corresponds to the specification of all the smeared
$n$-point functions of the quantum field. In particular, the
two-point function of the quantum field in the state $\omega$ is
defined by \cite{wald2} \be \label{2p-function-generic} \la
\hat{O}(\zeta)\hat{O}(\eta) \ra_{\omega}=
-\frac{\partial^{2}}{\partial s \partial t}\omega(\hat{W}(s\zeta +
t\eta))\, \exp(ist\w(\zeta , \eta)/2) \, \vert_{s=t=0} \, ,\ee
whereas the higher $n$-point functions of the quantum field are
defined similarly by using the Weyl relations. Since
$\hat{O}(\xi)=i({\cal{A}}(\bar{\xi})-{\cal{C}}(\xi))$, it is
straightforward to verify that the last definition gives
(\ref{2p-function}) for the algebraic state (\ref{magic}). In
terms of configuration and momentum operators the two-point
function (\ref{2p-function}) can be written explicitly as follows
(with $\zeta=(-g_{1},f_{1})$ and $\eta=(-g_{2},f_{2})$): \ba
\la \Phi^1_\zeta(\varphi) , \Phi^1_\eta(\varphi) \ra &=& \la
\Psi_{0},\varphi[f_{1}] \varphi[f_{2}]\Psi_{0}\ra + i\la
\Psi_{0},\varphi[f_{1}] \varphi[(1-iC)B^{-1}g_{2}]\Psi_{0}\ra  \nl
&+& i\la
\Psi_{0},\hat{\pi}[g_{1}]\varphi[(1-iC)B^{-1}g_{2}]\Psi_{0}\ra +
\la \Psi_{0},\hat{\pi}[g_{1}]\varphi[f_{2}]\Psi_{0}\ra  \, . \ea
Thus, the second moment of the measure $\d \mu$, that corresponds
to the choice $\zeta=\eta=(0,f)$, is given by \be \la
\Psi_{0},\varphi[f]^{2}\Psi_{0}\ra =\int_{\bar{\cal{C}}}\d \mu\,
\varphi[f]^{2}=\frac{1}{2}\int \d^3\!x\,fBf \, .\ee In general,
the $n$-th moment of the quantum measure goes as the $n/2$ power
of the covariance (two-point function of $\d \mu$) \cite{glimm} ,
$$\int_{\bar{\cal{C}}}\d \mu\, \varphi[f]^{n} = c_{n}
\biggl( \frac{1}{2}\int fBf \biggr)^{n/2}\, , $$ where $c_{n}$ is
a constant. Thus, as we expect from the previous discussion
(section \ref{sec:4}-\ref{sec:4.c}), the whole properties of the
measure will be dependent of those satisfied by $B$.

Let us now comment on the reverse question, namely how to find the
corresponding Fock state, given a state in the functional
representation. This problem is more involved. Let us recall what
happens in the case of a harmonic oscillator, In that case, the
``Fock representation" is given by the states expanded on the
basis of the form $\la n|\psi\ra$, and the corresponding image of
the basis states are the Hermite polynomials $H_n(q)$ in the
Schr\"odinger picture. Now, given a state $\psi(q)$, one can
always decompose it in the basis given by the Hermite polynomials
$\psi(q)=\sum_n a_nH_n(q)$, then the corresponding state in the
Fock picture is given by $|\psi\ra=\sum_n a_n|n\ra$. That is, the
state is given by the array of coefficients
$(a_0,a_1,\ldots,a_n,\ldots)$ in the (Hermite) expansion. In the
case of the field theory, given any Schr\"odinger state, one would
have to decompose it in a ``Hermite expansion" in order to find
its corresponding state in the Fock picture. We will not attempt
to do so in this paper. Note that this can indeed be done when the
representation is complex analytic, a la Bargmann, as done in
\cite{aa:am80}.

\section{Examples}
\label{sec:6}

We have stated  that there are infinite inequivalent
representations of the CCR \cite{wald2}. The structure responsible
for this fact can be taken as the complex structure $J$. In the
Fock representation, it is responsible for the separation of the
space of solutions into `positive' and `negative' frequency
solutions (to follow the standard terminology), which gives
meaning to the notion of particle. In the Schr\"odinger
representation the complex structure manifests itself in the
quantum measure and in the representation of the operators. The
conditions for when two different representations are equivalent
are well understood in both the Fock representation \cite{wald2}
and in the functional picture \cite{jero:phd}. A natural question
is then: given a fixed background spacetime, how do we choose the
{\em right} representation? This is indeed a highly non-trivial
question and there is no consensus on what the right answer should
be. For instance, some people argue that algebraic quantum field
theory tells us that the important object is the Weyl algebra and
not the representation itself \cite{wald2}. Other people still
think that the representation carries with it some important part
of the physics of the problem (see for instance \cite{helfer2}).

For the simplest system, namely Minkowski spacetime, the
quantization of the free scalar field was done in the early 20th
century without people worrying about the issue of uniqueness or
equivalence. This is the standard decomposition found in every
book on the subject. When QFT was constructed starting from
certain axioms \cite{cpt,glimm,baez} one of the main requirements
for the theory is that it be Poincar\'e invariant. On the other
hand, when QFT on curved spacetime was being formalized, it was
realized that a unique and natural choice of $J$ existed when the
background spacetime possesses a timelike Killing vector field
$t^a$ \cite{6}. When the spacetime is not stationary, the same
construction suggests the existence of a natural `time dependent'
complex structure. Naturally, Minkowski spacetime possesses a
sphere's worth of such vector fields (one for each global inertial
observer). A natural question one might ask is how the two
requirements are connected. That is, Can the requirement of
Poincar\'e invariance be compatible with the natural construction
from the Killing fields? Is there a unique choice satisfying both
conditions? As we shall see the answers are in the affirmative.

In this section we shall consider concrete examples of spacetimes
in which the quantum field theory is constructed. The remainder of
this  section has three parts. In the first one, we shall review
the standard construction of the quantum scalar field in Minkowski
spacetime, emphasizing the Poincar\'e invariance of the
representation coming from the time translation symmetries. In the
second part we shall consider stationary spacetimes and in the
last one, some simple cosmological solutions such as
Robertson-Walker spacetimes. Even when these examples have been
considered elsewhere, the approach taken here is novel and the
functional representation has not appeared before.

\subsection{Minkowski Spacetime}

Let us consider any two inertial observers $O_{1}$ and $O_{2}$ in
$(M,\eta_{ab})$. The coordinate system of each observer
corresponds to  spacetime foliations that define timelike Killing
vector fields $t_{1}^{a}$ and $t_{2}^{a}$, respectively. The
spacelike Cauchy surfaces
 $T_{t_{i}}(\Sigma)$ are chosen to be the (unique) normal to $t^{a}_{i}$
 (for $i=1,2$), namely the inertial frame in which the vector field
 $t^a_{i}$ is ``at rest''. The observers $O_{1}$ and $O_{2}$ decomposes
 solutions in its positive and negative frequency parts with respect to
 the associated Killing vector field, defining in such a way the
 ($\w$-compatible) complex structures
 $J_{1}=-(-{\pounds_{t_{1}}\pounds_{t_{1}}})^{-1/2}\pounds_{t_{1}}$
 and $J_{2}=-({-\pounds_{t_{2}}\pounds_{t_{2}}})^{-1/2}\pounds_{t_{2}}$.
  In general, every inertial observer defines a complex structure of the
  type $J=-({-\pounds_{\xi}\pounds_{\xi}})^{-1/2}\pounds_{\xi}$, where
  the vector field $\xi^{a}\in TM$ coincides with $(\partial /
  \partial t )^{a}$ in the observer coordinate system $X: M \to R^{4}$,
  $p \mapsto X(p)=(t,x,y,z)$.

Now, since $O_{1}$ and $O_{2}$ are inertial observers, they are
related by a Poincar\'e transformation $P$ which in turn induces a
transformation on the covariant phase space $\rm{\G_{S}}$. From
the one parameter family of embeddings $T_{t_{1}}$ we will choose
$T:=T_{t_{1}=0}$ to be the embedding from which the canonical
version of the theory will be constructed. The (active) Poincar\'e
transformation maps every Cauchy surface $T_{t_{1}}(\Sigma)$ onto
Cauchy surfaces $T_{t_{2}}(\Sigma)=P(T_{t_{1}}(\Sigma))$ that
corresponds to equal time surfaces for $O_{2}$. Thus, in
particular, we have that $T'(\Sigma)=P(T(\Sigma))$ and therefore
the corresponding Cauchy data associated to a solution $\phi$ of
the Klein-Gordon equation are different; namely, $\varphi
(x)=(\phi \circ T)(x)$ and $\pi (x)=(\sqrt{h}\pounds_{n}\phi \circ
T)(x)$ are the Cauchy data with respect to the embedding $T$,
whereas $\varphi' (x)=(\phi \circ T')(x)$
 and $\pi' (x)=(\sqrt{h'}\pounds_{n'}\phi \circ T')(x)$ are the
 corresponding Cauchy data with respect to $T'$. Hence, it is clear
 that the (active) Poincar\'e transformation $P:M\to M$ induces on the
 space of solutions, via $T'=P \circ T$, the transformation
 $\phi \mapsto \phi'=P^{*}\phi$. This symplectic transformation
 $P^{*}:\rm{\G_{S}} \to \rm{\G_{S}}$ in turn induces a ($\w$-compatible)
  complex structure $J'=P^{*}J_{1}P^{*-1}$. Indeed,
$$\mu_{1}(\phi , \phi)=\w(J_{1}\phi , \phi)=\w(P^{*}J_{1}P^{*-1}\phi',\phi')=:\mu'(\phi',\phi') \, .$$

Now, while on the one hand we have a natural complex structure
associated with the second observer,
 namely $J_{2}=-({-\pounds_{t_{2}}\pounds_{t_{2}}})^{-1/2}
\pounds_{t_{2}}$, on the other hand we have an induced complex
structure $J'=P^{*}J_{1}P^{*-1}$. Are these complex structures
related? Are they equivalent? It turns out that the answer is in
the affirmative; in fact these complex structures are exactly the
same. In order to see this, let us consider the Poincar\'e
transformation $P$ that induces a transformation $P_{T}$ such that
the image of the vector $t_{1}^{a}$ ``anchored'' on $p\in
T(\Sigma)$ is $t_{2}^{a}$ ``anchored'' on $p'=P\cdot p \in
T'(\Sigma)$. It is straightforward to verify (by using the mode
decomposition of the field) that the operator
$\Theta(t^{a}_{2}):=({-\pounds_{t_{2}}\pounds_{t_{2}}})^{-1/2}\pounds_{t_{2}}$
satisfies
$P^{*}\Theta(P_{T}^{-1}t^{a}_{2})P^{*-1}=\Theta(t_{2}^{a})$. Then,
\ba
-J_{2}=({-\pounds_{t_{2}}\pounds_{t_{2}}})^{-1/2}\pounds_{t_{2}}
&=&
({-\pounds_{P_{T}t_{1}}\pounds_{P_{T}t_{1}}})^{-1/2}\pounds_{P_{T}t_{1}}\nl
&=&
P^{*}({-\pounds_{t_{1}}\pounds_{t_{1}}})^{-1/2}\pounds_{t_{1}}P^{*-1}=-J'.
\nl  \ea That is to say, the natural complex structure associated
to an inertial frame is covariant under the Poincar\'e group.
Moreover, it is easy to see that given a positive (negative)
frequency solution $\phi^{+}$ ($\phi^{-}$) with respect to
$t_{1}^{a}$ and $J_1$, $J_{2}\phi^{+}=i\phi^{+}$
($J_{2}\phi^{-}=-i\phi^{-}$) and therefore $J_{1}\phi =
J_{2}\phi$. Since $J_{2}=P^{*}J_{1}P^{*-1}$, then
$(P^{*}J_{1}-J_{1}P^{*})\phi =0$; i.e., the natural complex
structure $J=-\Theta(t^{a})$ is invariant under Poincar\'e
transformations (furthermore, $J$ is unique since it satisfies
$\w(J\pounds_{t}\phi , \phi)=0$ \cite{6}) and consequently $P$
will be unitary{\footnote{We say that $J$ is invariant relative to
a given symplectic group if it commutes with all the symplectic
transformations in the group. If the complex structure is
invariant under a given symplectic transformation $T$ on
$\rm{\G_{S}}$, $T$ will be unitary relative to the complex
pre-Hilbert structure provided by (\ref{innp}) \cite{baez}.}}
relative to the complex pre-Hilbert structure $(\, \rm{\G_{S}}\, ,
\, \mu=\w \, \lceil \, J \,)$. To summarize, we have proved that
the usual positive-negative frequency decomposition found
everywhere is indeed the unique complex structure that is
Poincar\'{e} invariant.

The Poincar\'e invariant complex structure on $\rm{\G_{S}}$,
$J=-\Theta(t^{a})$, has a counterpart $J_{{\rm{\G}}}$ on the
Cauchy data space, provided by the (embedding-dependent)
isomorphism $I_{T}: {\rm{\G}} \to \rm{\G_{S}}$ through
$J_{{\rm{\G}}}=I_{T}^{-1}JI_{T}$. This relation and the general
form (\ref{est.com}) implies that \be \label{jcov-jcano} A\varphi
+ B\pi = -T^{*}[J \phi]\: , \:\:\:\:\: C\pi + D\varphi = -
T^{*}[\sqrt{h}\pounds_{n}(J \phi)] \, . \ee If $T$ represents an
``inertial embedding'' of $R^{3}$ into $M$, it is not difficult to
see that  the complex structure $J_{{\rm{\G}}}$
 is given by $J_{{\rm{\G}}}(\varphi,\pi) (-(-\Delta+m^2)^{-1/2}\pi,
 (-\Delta+m^2)^{1/2}\varphi)$,
 which means that $A=C=0$, $B=(-\Delta+m^2)^{-1/2}$ and
 $D=-(-\Delta+m^2)^{1/2}$. The quantum measure is then
 $``\d\mu=e^{-\int\varphi(-\Delta+m^2)^{1/2}\varphi}\;{\cal D}\varphi"$.
 Thus, we obtain the Gaussian measure and Fock representations existing
 in the literature \cite{glimm,baez,veli}. As should be clear from the
 discussion, this represents a very particular case of the general
 formulae presented above. The momentum operator (\ref{repdepi-conC})
  is given by
\begin{equation}
\label{repdepi-minkowski} (\hat{\pi}[g]\cdot
\Psi)[\varphi]=-i\int_{R^{3}}\d^3\!x \biggl(g
{{\delta}\over{\delta \varphi}} -\varphi(-\Delta+m^2)^{1/2}g
\biggr) \Psi[\varphi] \, ,
\end{equation}
and provides us with a preferred representation in the sense that
it arose by requiring Poincar\'e invariance.

\subsection{Stationary Spacetimes}

We shall consider here spacetimes which possess a timelike Killing
vector field $t^a$ and such that a spacelike hypersurface,
orthogonal to the orbits of the isometry, might not exist. In
particular, the Killing vector field provides us with a natural
slicing of spacetime into space and time, and the shift function
necessarily will be different from zero, when the vector field is
not hypersurface orthogonal.

Now, it is well-known \cite{6,7,3women} that given a real Hilbert
space ${\H}$ and a non degenerate symplectic structure defined
thereon, then there exists a complex structure $J$ on ${\H}$ and a
real scalar product $\mu$ such that $\w(x,y)=\mu(Jx,y)$. Indeed,
by the Riesz lemma there exists a continuous linear operator
$E:{\H}\to{\H}$ such that $\w(x,y)=\la Ex,y\ra$, where $\la \,
\cdot , \cdot \, \ra$ denotes the given scalar product on ${\H}$.
Since the complex structure is skew, then $E^{*}=-E$ and
$-E^{2}\geq 0$. It is not difficult to see that $-E^{2}$ is a
continuous linear operator and therefore there is (by the square
root lemma \cite{reeds}) a unique bounded operator $\vert E \vert$
with $\vert E \vert \geq 0$ and $\vert E \vert = \sqrt{-E^{2}}$.
Moreover, it can be shown that $\vert E \vert$ is self adjoint
injective, has a dense range and an inverse $ \vert E \vert
^{-1}$. From the polar decomposition theorem \cite{reeds}, since
$E$ is a bounded linear operator on ${\H}$, there is a partial
isometry $U$ such that $\vert E \vert U = E$. This partial
isometry defines a complex structure $J=\vert E \vert ^{-1}E$:
$-E^{2}=\vert E \vert^{2}$ implies $J^{2}=-{\bf{1}}$ and is
orthogonal $J^{*}=-J=J^{-1}$. Hence, the scalar product $\mu$ on
${\H}$ is defined by
\[
\mu(x,y)=\la \vert E \vert x , y\ra \, .
\]

Let $F$ be the operator that dictates classical evolution in the
canonical framework. The idea is to prescribe the inner product on
the Cauchy data space $\G$ as $\la F \, \cdot\,  , \, \cdot \, \ra
= \w (\, \cdot \, , \, \cdot \,)$, in such a way that $F$ will be
an
 anti-self adjoint operator and hence $-F^{2}$ will be non negative.
 From the Klein-Gordon Hamiltonian
\be \label{hamil-kg} H_{KG}=\frac{1}{2}\int_{\Sigma} \d^{3}\!x \,
\sqrt{h}\,\biggl(\frac{N}{h} \pi^{2}+Nh^{ab}D_{a}\varphi
D_{b}\varphi + Nm^{2}\varphi^{2}+ \frac{2\pi}{\sqrt{h}}
N^{a}D_{a}\varphi \biggr)\, , \ee it is easy to see that the
equations of motion are given by
\begin{eqnarray}
\label{matrix-form} \left( \begin{array}{c} \dot{\varphi} \\
\dot{\pi} \end{array} \right) = F\left( \begin{array}{c} \varphi \\
\pi \end{array} \right)\: , \:\: \mbox{ where }
 F=\left( \begin{array}{cc} N^{a}D_{a} & \ld \\ -\Theta & \Lambda
 \end{array} \right)
\end{eqnarray}
and $\Theta := -\sqrt{h}(N D^{a}D_{a}+ D^{a}ND_{a}-Nm^{2})$,
$\Lambda :=\sqrt{h}D_{a}({\ld}^{a})+N^{a}D_{a}) $, $\ld
:=N/\sqrt{h}$,
 ${\ld}^{a}:=N^{a}/\sqrt{h}$. From
 $\w((^{\varphi}_{\pi})_{1},F^{*}(^{\varphi}_{\pi})_{2})=
 \w(F(^{\varphi}_{\pi})_{1},(^{\varphi}_{\pi})_{2})$ it is
 clear that $F^{*}=-F$ indeed, and consequently there is a unique
  $\vert F \vert$, non negative, that satisfies
\be \label{sqrt-eq} \vert F \vert \vert F \vert = -F^{2} \, . \ee
Let $\vert F \vert$ be the matrix
\begin{eqnarray}\label{square-root}
\vert F \vert = \left( \begin{array}{cc} a & b \\ c & d
\end{array} \right) \, ,
\end{eqnarray}
since $\vert F \vert$ is self adjoint, then the operators $a$,
$b$, $c$ and $d$ must be such that \be \int_{\Sigma}\pi a \varphi
= \int_{\Sigma}\varphi d \pi \: , \:\: \int_{\Sigma} \varphi_{1} c
\varphi_{2} = - \int_{\Sigma} \varphi_{2} c \varphi_{1} \: , \:\:
\int_{\Sigma} \pi_{1} b \pi_{2} = - \int_{\Sigma}
 \pi_{2} b \pi_{1} \, .
\ee Now, the relationship (\ref{sqrt-eq}) implies the following
system of operator equations that one has to solve in order to
find the general expression for the complex structure $J$, \be
\label{abc-eq} a^{2}+bc=\ld \Theta -N^{a}D_{a}N^{b}D_{b} \, .\ee
\be \label{abd-eq} ab+bd=-(N^{a}D_{a}\ld +\ld \Lambda) \, .\ee \be
\label{acd-eq} ca+dc=\Theta N^{a}D_{a}+\Lambda \Theta \, .\ee \be
\label{cbd-eq} cb+d^{2}=\Theta \ld - \Lambda^{2} \, .\ee

Note that if we restrict our attention to orthogonal foliations
(which in turn are natural for static spacetimes) then we recover
the particular case specified in the appendix of \cite{6} by
Ashtekar
 and Magnon (with slight differences that come from the fact that,
 instead of scalars, here the momenta are scalars densities of weight one).
 Indeed, if $N^{a}=0$, then (\ref{abc-eq})-(\ref{cbd-eq}) yields
\be \label{shift-zero} a^{2}+bc=\ld \Theta \:\: , \:\:
ab+bd=ca+dc=0 \:\: , \:\: cb+d^{2}=\Theta \ld \, ,\ee which means
that $a=(\ld \Theta)^{1/2}$, $d=(\Theta \ld)^{1/2}$, and $b=c=0$.
Therefore
\[
\vert F \vert ^{-1}=\left( \begin{array}{cc} (\ld \Theta)^{-1/2} &
0 \\ 0 & (\Theta \ld)^{-1/2} \end{array} \right)
\]
 and the complex structure that one obtains by virtue of the polar
 decomposition is
\be \label{cs-shift-zero} J=\left( \begin{array}{cc} 0 &
\Theta^{-1/2} {\ld}^{1/2} \\ - {\ld}^{-1/2}\Theta^{1/2} & 0
\end{array} \right) \, , \ee that is to say $B=\Theta^{-1/2}
{\ld}^{1/2}$, $D=-{\ld}^{-1/2} \Theta^{1/2}$ and $A=C=0$ (c.f.
eq.(\ref{est.com})).

Thus, the requirements in the appendix of \cite{6} are equivalent
to perform the polar decomposition by using the infinitesimal
evolution operator associated to the Hamiltonian density
$${\H}_{KG}=\frac{N \sqrt{h}}{2}\,\biggl(\frac{\pi^{2}}{h}+D^{a}
\varphi D_{a}\varphi + m^{2}\varphi^{2}\biggr) \, .$$

In this particular case, the form of the measure is
\[
``\d\mu=e^{-\int\varphi(\ld^{-1/2}\, \Theta^{1/2}\varphi)}\;{\cal
D}\varphi" \, ,
\]
Let us end our detour into orthogonal foliations and return to the
more general case where $b$ and $c$ will be different from zero,
and hence with non zero shift (i.e., $N^{a}\neq 0$). If this is
the case, the operators that one obtains via the polar
decomposition, and that specify the complex structure
(\ref{est.com}), are given by \ba
 A=\Upsilon N^{a}D_{a}-\Delta
\Theta  \:\: , \:\:\:\: B= \Upsilon \ld +\Delta \Lambda \:\: ,\nl
C=-d^{-1}c\Upsilon \ld -b^{-1}a\Delta \Lambda \:\:
 , \:\:\:\: D=b^{-1}a\Delta \Theta -d^{-1}c\Upsilon N^{a}D_{a}\, ,
\ea where $\Delta :=(c-db^{-1}a)^{-1}$ and $\Upsilon
:=(a-bd^{-1}c)^{-1}$. This complex structure on the Cauchy data
space $\G$ corresponds to the
 (negative of the) complex structure induced by the covariant one
 $J=({-\pounds_{t}
 \pounds_{t}})^{-1/2}\pounds_{t}$ that is the unique complex structure
 that satisfies the energy-requirement, as Ashtekar and Magnon showed
 \cite{6,aa:am80}. More precisely, let $T_{t}$ be the uniparametric family
 of embeddings of $\Sigma$ into $M$ that corresponds to a slicing with
  respect to the timelike Killing vector field $t^{a}$, then the induced
   complex structure on $\G$ is given by $I^{-1}_{T}
   (-{\pounds_{t}\pounds_{t}})^{-1/2}\pounds_{t}I_{T}$, where $I_{T}:
    {\rm \G \to \G_{S}}$ is the natural isomorphism (associated to an
     embedding $T$ of the family) between the space of Cauchy data and
     the space of solutions. Thus, $F=I^{-1}_{T}\pounds_{t}I_{T}$ and
     $\vert F \vert^{-1} =I^{-1}_{T}({-\pounds_{t}\pounds_{t}})^{-1/2}I_{T}$.

The quantum measure is then
$$``\d\mu=e^{-\int\varphi(\Upsilon \ld +\Delta
\Lambda)^{-1}\varphi}\;{\cal D}\varphi" \, ,$$ and the preferred
representation for the momentum operator (\ref{repdepi-conC}),
selected by the energy-requirement, is given by
\begin{equation}
\label{repdepi-generic} (\hat{\pi}[g]\cdot
\Psi)[\varphi]=-i\int_{\Sigma}\d^3\!x \biggl(g
{{\delta}\over{\delta \varphi}} -\varphi [1+i(d^{-1}c\Upsilon \ld
+b^{-1}a\Delta \Lambda)](\Upsilon \ld +\Delta \Lambda)^{-1}g
\biggr) \Psi[\varphi] \, .
\end{equation}

\subsection{Cosmological Solutions}

Let us consider now the Robertson-Walker cosmological model of
homogeneous and isotropic universes. That is to say, spacetimes
which spatial geometry, restricted by the homogeneity and isotropy
requirements, is that one of (a) a sphere, (b) flat Euclidean
space, or (c) a hyperboloid. For this class of
 spacetimes, the four-dimensional metric $g_{ab}$ may be expressed as
 $g_{ab}=-u_{a}u_{b}+h_{ab}$,
where $u^{a}$ are the tangents to the world line of isotropic
observers, which are orthogonal to the $3-d$ homogeneous surfaces
with metric $h_{ab}$. Thus, the line element for the three
possible spatial geometries, described in spherical, Cartesian and
hyperbolic coordinates, respectively, is given by
\begin{displaymath}
\label{rob-walk} \d s^{2}=-\d\tau^{2}+a^{2}(\tau) \left\{
\begin{array}{ll}\d\psi^{2}+
\sin^{2}\psi(\d\theta^{2}+\sin^{2}\theta \d\phi^{2})\, , & 0\leq
\psi
\leq 2\pi \\
\d x^{2}+\d y^{2}+\d z^{2} &  \\
\d\psi^{2}+\sinh^{2}\psi(\d\theta^{2}+\sin^{2} \theta
\d\phi^{2})\, , & 0\leq \psi < \infty  \end{array}\right.
\end{displaymath} where $a(\tau)$ is a positive function of the
proper time $\tau$ of any of the isotropic observers.

Now, the world lines of the isotropic observers define a time-like
hypersurface orthogonal unit vector field $(\partial /
\partial \tau)^{a}$, which in turns determines canonically a
complex structure of the form (\ref{cs-shift-zero}) on the space
of Cauchy data by virtue of the polar decomposition of
$\pounds_{\tau}$. Since the lapse function is equal to one, then
$\Theta = -\sqrt{h}(D^{a}D_{a}-m^{2})$ and hence the
 complex structure (\ref{cs-shift-zero}) will be specified, for each one
 of the possible geometries, by
$$(D^{a}D_{a},\ld)=$$
\begin{displaymath}
\label{rob-walk-cs} \left\{
\begin{array}{ll}\biggl(\frac{1}{a^{2}}
(\partial^{2}_{\psi}+2\cot\psi
\partial_{\psi})+\frac{1}{a^{2}\sin^{2}\psi}
(\partial^{2}_{\theta}+\cot\theta
\partial_{\theta}+\frac{1}{\sin^{2}\theta}
\partial^{2}_{\phi}) \:\: , \: 1/a^{3}\sin^{2}\psi\sin\theta \biggr)
\\ \biggl( \frac{1}{a^{2}}(\partial^{2}_{x}+\partial^{2}_{y}+\partial^{2}_{z})
\: \: , \: 1/a^{3}\biggr)
  \\  \biggl(\frac{1}{a^{2}}(\partial^{2}_{\psi}+2\coth\psi \partial_{\psi})
  +\frac{1}{a^{2}\sinh^{2}\psi}(\partial^{2}_{\theta}+\cot\theta
  \partial_{\theta}+\frac{1}{\sin^{2}\theta}\partial^{2}_{\phi}) \:\: ,
  \: 1/a^{3}\sinh^{2}\psi\sin\theta \biggr)  \end{array}\right.
  \end{displaymath}
The presence of $a(\tau)$ in the complex structure, through
$D^{a}D_{a}$ and $\ld$, implies that it is a time-dependent
complex structure for each
 (dust or radiation-filled) Robertson-Walker universe and hence the vacuum
 state is unstable. Indeed, let $T$ be the embedding (of $\Sigma$ into the
 spacetime) that corresponds to the ``equal time'' Cauchy surface
 $\tau=\tau_{1}$, and let $J_{i}$ be the complex structure defined by
 the pair $(D^{a}D_{a},\ld)$ at $\tau = \tau_{i}$. Then, the covariant
  complex structure is given by $J=I_{T}J_{1}I_{T}^{-1}$, where
  $I_{T}:{\rm \G \to \G_{S}}$ is the natural isomorphism associated
  to $T$. The vacuum expectation associated to the fictitious complex
  structure $J'=I_{T}J_{2}I_{T}^{-1}$ is, in general, different from
  the vacuum expectation associated to $J$. The algebraic states certainly
  will be different functionals on the Weyl algebra and consequently the
  Fock spaces obtained via the GNS construction disagree.

Finally,
 let us consider the particular case of a dust-filled Robertson-Walker
flat universe. The explicit form of the complex structure in this
background
 is given by  \be
\label{cs-rw-flat} J=\left( \begin{array}{cc} 0 & \frac{1}{a^{2}}
(-\Delta+M^{2})^{-1/2}
\\ -a^{2}(-\Delta+M^{2})^{1/2} & 0 \end{array} \right) \, ,
\ee where $\Delta$ denotes the Laplacian operator in Cartesian
coordinates,
 $a(\tau)=(9C/4)^{1/3}\tau^{2/3}$ (with $C=8\pi \rho a^{3}/3$ a constant)
  and $M(\tau):=ma(\tau)$. Thus, the Gaussian measure and the momentum
  operator are
$$
``\d\mu=e^{-\int\d^3\!x\,\varphi a^{2}(-\Delta+M^{2})^{1/2}
\varphi}\;{\cal D} \varphi" \, , $$ and
$$(\hat{\pi}[g]\cdot
\Psi)[\varphi]=-i\int_{\Sigma}\d^3\!x\, \biggl(g
{{\delta}\over{\delta \varphi}} -\varphi
a^{2}(-\Delta+M^{2})^{1/2} g \biggr) \Psi[\varphi] \, . $$ The
static case $a=$constant is, of course, the case of Minkowski
space (c.f. eq(\ref{repdepi-minkowski})).

\section{Discussion}
\label{sec:7}

The aim of this paper was to present in a self-contained manner
the main steps necessary to construct the Fock and Schr\"odinger
representations for a real Klein-Gordon field on Minkowski
spacetime, for arbitrary embeddings of the ``constant time
hypersurface". We first briefly reviewed the canonical
quantization procedure, emphasizing the definitions of the
classical objects such as phase space and observables of a scalar
field which are needed for its quantization.

For the covariant Fock quantization we started from the phase
space of classical solutions of the Klein-Gordon equation and
constructed the corresponding one-particle Hilbert space. In this
construction we emphasized the role of the complex structure $J$
as responsible for the infinite freedom in the choice of the
quantum representation for the Fock Hilbert space (alternatively
one can use the metric $\mu$ \cite{wald2}).

In the functional Schr\"odinger representation we discussed in
detail the role played by certain classical constructs. For
instance we have, in addition to $J$, a second classical object,
namely the choice of embedding $T$. We argued that it is necessary
to consider a probabilistic measure instead of the naive Lebesgue
measure which, in fact, does not exist in this case. We showed
that the choice of complex structure $J$ and embedding $T$ was
manifest at the level of the measure $\d\mu$ and in the
representation of the momentum operator. In connection with the
Hilbert space needed for the functional representation we made the
distinction between the classical ${\cal C}$ and the quantum
$\overline{\cal C}$ configuration spaces, but we have not
specified $\overline{\cal C}$.\footnote{Recall that in the case of
Minkowski spacetime and for  flat embeddings, where $\Sigma$ is a
Euclidean space, the quantum configuration space is the space of
tempered distributions ${\cal J}^*$  on $\Sigma$. However, in
order to define this space one uses the linear and Euclidean
structure of $\Sigma$ and its generalization to curved manifolds
is not trivial.} This problem lies beyond the scope of this paper,
and shall be reported elsewhere \cite{ccq:qcs}. In order to relate
explicitly the Fock and Schr\"odinger representations we have used
the algebraic formulation of quantum field theory, by means of the
GNS construction.  Finally, we have discussed the explicit mapping
that relates the states in both representations, and provided
several examples of such constructions.

In this paper, we have only considered the Schr\"odinger
functional picture where states are functions of the (real)
configuration variable $\varphi$, or as sometimes it is known, the
`vertical polarization'. There are alternative {\it functional
representations} that have been considered in the literature (see
for instance Ref.\cite{aa:am80}). This representation are however,
the functional equivalence of the Bargmann representation in
quantum mechanics, where states are holomorphic functions on phase
space (i.e. the holomorphic polarization), with complex
coordinates $z^i$. This representation is much closer to the Fock
representation as is clear from quantum mechanics where the basis
states $|n_j\rangle$ are given by the functions $\langle
z_i|n_j\rangle=z_i^{n_j}$. We shall leave the comparison between
these two functional description for a future publication.

\section*{Acknowledgments}

We would like to thank A. Ashtekar for discussions, R. Jackiw for
drawing our attention to Refs.~\cite{reviews,jackiw} and J.
Velhinho for comments. This work was in part supported by
DGAPA-UNAM grant No. IN112401, by CONACyT grants J32754-E and
36581-E, NSF grant No. PHY-0010061, and US DOE grant
DE-FG03-91ER40674.
 J.C. was supported by a UNAM
(DGEP)-CONACyT Graduate Fellowship. H.Q. thanks UC-MEXUS for
support.

\appendix

\section{The example of the Harmonic Oscillator}
\label{app:a}

Let us consider in this section the example of a simple harmonic
oscillator in $N$ dimensions. The configuration space is ${{\cal
C}}=\{ q^i\}$, and the phase space $\Gamma=(q^i,p_i)$. The
symplectic structure is given by
\[
\Omega_{ab}=2\nabla_{[a}p_i\nabla_{b]}q^i \, ,
\]
such that $\{q^i,p_j\} =\delta^i_j$. The standard complex
structure $J$ compatible with the symplectic two form is
 \be
{J^a}_b=\left(\frac{\partial}{\partial p_i}\right)^a\, \nabla_b
q^i-\left(\frac{\partial}{\partial q^i}\right)^a\, \nabla_b p_i \,
.
 \ee
It is straightforward to see that the induced metric
$\mu_{ab}:{J^c}_a\Omega_{cb}$ is given by \be
 \mu_{ab}=\nabla_a q^i\nabla_b q_i+\nabla_a p^i\nabla_b p_i \, .
\ee The complex structure then has the following action on the
basis vectors on $T\Gamma$, \be
J\cdot\left(\frac{\partial}{\partial
q^i}\right)=\left(\frac{\partial}{\partial p_i}\right)\qquad{\rm
and}\qquad J\cdot\left(\frac{\partial}{\partial
p_i}\right)=-\left(\frac{\partial}{\partial q^i}\right) \, .
 \ee
Let us now consider the observables. Now the vector indicated by
$Y^a=(q^i,p_j)^a$ is given by,
\[
Y^a=q^i \left(\frac{\partial}{\partial q^i}\right)^a + p_i
\left(\frac{\partial}{\partial p_i}\right)^a \] and the one-form
$\lambda_a=(-\alpha_i ,-\beta^j)_a$ is given by,
\[
\lambda_a=-\alpha_i(\nabla_a q^i)-\beta^i(\nabla_ap_i)
\]
such that \[F_\lambda(Y):=-\lambda_aY^a=\alpha_iq^i+\beta^ip_i.\]
Recall that there is also a label vector given by
$(-\beta,\alpha)^a$, that gives rise to particular configuration
and momentum observables as follows
\[
Q(\alpha_i):=F_{(-\alpha,0)}(Y)=\w_{ab}(0,\alpha_{i})^{a}(q^k,p_{j})^{b}
=\alpha_iq^i
\]
and
\[
P(\beta^i):=F_{(0,-\beta)}(Y)=\w_{ab}(-\beta^i,0)^{a}(q^k,p_{j})^{b}
=\beta^ip_i
\]
Let us suppose that we have found a representation of the basic
observables as operators, namely we have $\hat{q}^i$ and
$\hat{p}_i$ acting on a Hilbert space $\H$. The corresponding
operator is given by $\hat{\cal
O}[(-\beta,\alpha)]:=\alpha_i\hat{q}^i+\beta^i\hat{p}_i$.
 Let us now construct
the creation and annihilation operators for a general observable:
\be {\cal C}[(-\beta,\alpha)]:=\frac{1}{2\hbar}\left[ \hat{\cal
O}[J\cdot(-\beta,\alpha)]+i \hat{\cal O}[(-\beta,\alpha)]\right]=
\frac{1}{2\hbar}\left[(-\beta_i+i\alpha_i)(\hat{q}^i-i\hat{p}^i)\right]
\ee
 and for the annihilation operator we have,
\be {\cal A}[(-\beta,\alpha)]:=\frac{1}{2\hbar}\left[ \hat{\cal
O}[J\cdot(-\beta,\alpha)]-i \hat{\cal O}[(-\beta,\alpha)]\right]=
\frac{1}{2\hbar}\left[(-\beta_i-i\alpha_i)(\hat{q}^i+i\hat{p}^i)\right]
\ee Let us note that if we define the operators, \be
\hat{a}^{i\dagger}:=\frac{1}{\sqrt{2\hbar}}(\hat{q}^i-i\hat{p}^i)
\ee and, \be
\hat{a}^{i}:=\frac{1}{\sqrt{2\hbar}}(\hat{q}^i+i\hat{p}^i) \ee and
the complex coordinates
$\zeta^i:=\frac{1}{\sqrt{2\hbar}}(-\beta^i+i\alpha^i)$, then we
have \be {\cal
C}[(-\beta,\alpha)]=\zeta_i\,\hat{a}^{i\dagger}\qquad {\rm
and}\qquad {\cal
A}[(-\beta,\alpha)]=\overline{\zeta}_i\,\hat{a}^{i} \ee Thus, we
get the standard relations between the ordinary Schr\"{o}dinger
and the oscillator representation, making also contact between the
notation we have used of defining the creation of a state labelled
by a `label vector' $\zeta$ and the ordinary creation and
annihilation operators used elsewhere.

\section{Non-trivial Vacua}
\label{app:b}

In Sec.~\ref{sec:4.a}, we discussed the possibility of defining a
Schr\"odinger representation in which the measure is the
``homogeneous" one. As mentioned before, this representation does
not exist from a rigorous viewpoint since the homogeneous measure
is not well defined \cite{draft}. However, one can ignore this and
pretend that this representation exists. We expect that in analogy
with the harmonic oscillator,  in this case the vacuum will have a
Gaussian form. However, there is no a-priori expression for it.
The purpose of this appendix is to develop this reasoning, and
construct then the non-trivial vacua associated to the Gaussian
measures. We know from the general discussion in
Sec.~\ref{sec:4.a}, that the ``momentum operator'' associated to
an homogeneous measure is represented as follows ($\hbar=1$):
\begin{equation}
\hat{\pi}[g]=-i \int \d^3\!x\; g(x){{\delta}\over{\delta
\varphi(x)}} \, .
\end{equation}
Now, applying the equation for the vacuum $\Psi_0$, namely ${\cal
A}(\zeta)\cdot\Psi_0=0$ for all $\zeta\in\Gamma$, we get from
(\ref{aniquil}) that (with $\hbar =1$) \be {\cal
A}(-g,f)\cdot\Psi_0= \frac{1}{2} \int_\Sigma\left(\varphi
[Dg-Cf-if]+[iAg-g-iBf]\frac{\delta}{\delta\varphi} \right)\Psi_0
=0 \, .\ee Let $\Lambda$ be such that
$\delta\Psi_0[\varphi]/\delta\varphi=\Lambda\Psi_0[\varphi]$.
Then, \be \int_\Sigma\Bigl( \varphi [Dg-Cf-if]+\Lambda [iAg-g-iBf]
\Bigr)\Psi_0 =0 \ee for all $(-g,f)\in\Gamma$. Given that $g$ and
$f$ are independent, the last equation should be valid for all
vectors of the type $(0,f)\in\Gamma$. Thus, using the first and
last relation in (\ref{relaciones2}) we have that \be \int_\Sigma
\Bigl(f(iB\Lambda +[i-A]\varphi)\Bigr)\Psi_0=0; \; \forall f \ee
which implies that $({\bf{1}}+iA)\varphi+B\Lambda=0$, then
$\Lambda=-B^{-1}\varphi-iB^{-1}A\varphi =
-({\bf{1}}-iC)B^{-1}\varphi$ (where we have used that
$B^{-1}A=-CB^{-1}$). We can now verify that, for all $g$, \be
\int_{\Sigma}\Bigl( \varphi Dg+\Lambda [iAg-g]
\Bigr)\Psi_0=\int_{\Sigma} g\Bigl(D\varphi-iC\Lambda- \Lambda
\Bigr)\Psi_0 \ee vanishes after substituting the expression for
$\Lambda$ and using the conditions satisfied by the operators
$A,B,C,D$. Thus, we can conclude that the condition in the vacuum
reads,
\begin{equation}
\frac{\delta\Psi_0[\varphi]}{\delta\varphi}=-[({\bf{1}}-iC)B^{-1}\varphi]
\Psi_0[\varphi]=:-({\cal
Q}\cdot\varphi)\Psi_0[\varphi],\label{vaccon}
\end{equation}
where we have defined the operator ${\cal
Q}:=({\bf{1}}-iC)B^{-1}$. We make then the ansatz,
\begin{equation}
\Psi_0[\varphi]=e^{\alpha\int_\Sigma\varphi{\cal Q}\cdot\varphi}
\, .
\end{equation}
Let us now show that this state indeed satisfies (\ref{vaccon}).
Let $\{\varphi_\lambda\}$ be a one parameter family of field
configurations and
$\delta\varphi:=\d\varphi_\lambda/\d\lambda|_{\lambda=0}$, then,
\begin{equation}
\frac{\d\Psi_0}{\d\lambda}=\alpha\int_\Sigma\left[
\frac{\d\varphi_\lambda}{\d\lambda}({\cal
Q}\cdot\varphi_\lambda)+\varphi_\lambda\left({\cal
Q}\cdot\frac{\d\varphi_\lambda}{\d\lambda}\right)\right]\Psi_0=\alpha
\int_{\Sigma}[\tilde\varphi_\lambda{\cal Q}\cdot\varphi_\lambda+
\varphi_\lambda{\cal Q}\cdot\tilde{\varphi}_\lambda]\Psi_0
\end{equation}
where $\tilde\varphi_\lambda:=\d\varphi_\lambda/\d\lambda$. Let us
consider the term of the form $\int g{\cal Q}g'$, for all $g$ and
$g'$ in $C^\infty_0(\Sigma)$. Since $B$ is symmetric, $B^{-1}$
will also be, and then $\int g{\cal Q}g'=\int g'B^{-1}g-i\int
gC(B^{-1}g')$, but since $\int gC(B^{-1}g')=-\int
Ag(B^{-1}g')=-\int g'B^{-1}Ag$, using the identity
$B^{-1}A=-CB^{-1}$ we have that $\int gCB^{-1}g'=\int g'CB^{-1}g$.
Therefore, ${\cal Q}$ is symmetric. Hence, we can conclude that,
\begin{equation}
\left.\frac{\d\Psi_0}{\d\lambda}\right|_{\lambda=0} =\int_\Sigma
\tilde{\varphi}_\lambda (2\alpha {\cal
Q}\cdot\varphi_\lambda)\Psi_0|_{\lambda=0}=\int
\delta\varphi\left[\frac{\delta\Psi_0[\varphi]}{\delta\varphi}\right]
\, ,
\end{equation}
which implies that $\delta\Psi_0[\varphi]/\delta\varphi=2\alpha
({\cal Q}\cdot\varphi)\Psi_0[\varphi]$. Therefore, $\alpha=-1/2$
and the vacuum, in the ``homogeneous" Schr\"odinger
representation, is given by the functional,
\begin{equation}
\Psi_0[\varphi]=e^{-\frac{1}{2}\int_\Sigma\varphi(B^{-1}-iCB^{-1})
\varphi} \, .
\end{equation}
Thus, if we were to absorb the vacuum into the measure, we would
have $``\d\mu={\cal D}\varphi\;\overline{\Psi_0}\Psi_0={\cal
D}\varphi \;e^{-\int_{\Sigma}\varphi B^{-1}\varphi}"$ which is
precisely the Gaussian measure given by (\ref{medida}).

\end{document}